\documentclass[preprintnumbers,article,amsmath,amssymb,floatfix,10pt,prd,twocolumn,
superscriptaddress,nofootinbib]{revtex4-2}
\usepackage{bm}
\usepackage{amsfonts}
\usepackage{latexsym}
\usepackage[latin1]{inputenc}
\usepackage{graphicx}
\usepackage{amsmath}
\usepackage{bigints} 
\usepackage{palatino}
\usepackage{mathpazo}
\usepackage{textcomp}
\usepackage{float}
\usepackage{booktabs}
\usepackage{dcolumn}
\usepackage{ragged2e}
\usepackage{hyperref}
\usepackage{relsize} 
\hypersetup{colorlinks,citecolor=blue}
\hypersetup{colorlinks=true,linkcolor=red,filecolor=magenta,    urlcolor=blue}
\usepackage{amsmath}
\usepackage{xcolor}
\usepackage{orcidlink}
\usepackage{epsfig}
\usepackage{subfigure}
\usepackage{commath}

\DeclareFontFamily{U}{BOONDOX-calo}{\skewchar\font=45 }
\DeclareFontShape{U}{BOONDOX-calo}{m}{n}{
  <-> s*[1.05] BOONDOX-r-calo}{}
\DeclareFontShape{U}{BOONDOX-calo}{b}{n}{
  <-> s*[1.05] BOONDOX-b-calo}{}
\DeclareMathAlphabet{\mathcalboondox}{U}{BOONDOX-calo}{m}{n}
\SetMathAlphabet{\mathcalboondox}{bold}{U}{BOONDOX-calo}{b}{n}
\DeclareMathAlphabet{\mathbcalboondox}{U}{BOONDOX-calo}{b}{n}

\def\jnl@style{\it}
\def\aaref@jnl#1{{\jnl@style#1}}

\def\aaref@jnl#1{{\jnl@style#1}}

\def\aj{\aaref@jnl{AJ}}                   
\def\apj{\aaref@jnl{ApJ}}                 
\def\apjl{\aaref@jnl{ApJ}}                
\def\apjs{\aaref@jnl{ApJS}}               
\def\apss{\aaref@jnl{Ap\&SS}}             
\def\aap{\aaref@jnl{A\&A}}                
\def\aapr{\aaref@jnl{A\&A~Rev.}}          
\def\aaps{\aaref@jnl{A\&AS}}              
\def\mnras{\aaref@jnl{Mon.~Not.~Roy.~Astron.~Soc.}}             
\def\prd{\aaref@jnl{Phys.~Rev.~D}}        
\def\prc{\aaref@jnl{Phys.~Rev.~C}}  
\def\prl{\aaref@jnl{Phys.~Rev.~Lett.}}    
\def\qjras{\aaref@jnl{QJRAS}}             
\def\skytel{\aaref@jnl{S\&T}}             
\def\ssr{\aaref@jnl{Space~Sci.~Rev.}}     
\def\zap{\aaref@jnl{ZAp}}                 
\def\nat{\aaref@jnl{Nature}}              
\def\aplett{\aaref@jnl{Astrophys.~Lett.}} 
\def\apspr{\aaref@jnl{Astrophys.~Space~Phys.~Res.}} 
\def\physrep{\aaref@jnl{Phys.~Rep.}}      
\def\physscr{\aaref@jnl{Phys.~Scr}}       
\def\commat{\aaref@jnl{Comm.~Math.~Phys.}}              
\def\science{\aaref@jnl{Science}}               
\def\cqg{\aaref@jnl{Classical Quant.~Grav.}}            
\def\jpcs{\aaref@jnl{JPCS}}                                     
\def\ijmpd{\aaref@jnl{Int.~J.~Mod.~Phys.~D}}                    
\def\grg{\aaref@jnl{Gen.~Relat.~Gravit.}}               
\def\rpp{\aaref@jnl{Rep.~Prog.~Phys.}}          
\def\npa{\aaref@jnl{Nucl.~Phys.~A}}        
\def\lrr{\aaref@jnl{Living Rev.~Rel.}}                   
\def\jcap{\aaref@jnl{J.~Cosmology Astropart.~Phys.}}    
\def\rmp{\aaref@jnl{Rev.~Mod.~Phys.}}   
\def\epjc{\aaref@jnl{Eur.~Phys.~J.~C}} 
\def\plb{\aaref@jnl{~Phy.~Lett.~B}} 
\def\mpla{\aaref@jnl{Mod.~Phy.~Lett.~A}} 
\def\arxiv{\aaref@jnl{arxiv.org}}

\allowdisplaybreaks[1]

\begin{document}
\color{black} 
\title{Conformally symmetric wormhole solutions supported by non-commutative geometries in the context of $f(Q, T)$ gravity}

\author{Moreshwar Tayde\orcidlink{0000-0002-3110-3411}}
\email{moreshwartayde@gmail.com}
\affiliation{Department of Mathematics, Birla Institute of Technology and
Science-Pilani,\\ Hyderabad Campus, Hyderabad-500078, India.}

\author{Zinnat Hassan\orcidlink{0000-0002-6608-2075}}
\email{zinnathassan980@gmail.com}
\affiliation{Department of Mathematics, Birla Institute of Technology and
Science-Pilani,\\ Hyderabad Campus, Hyderabad-500078, India.}

\author{P.K. Sahoo\orcidlink{0000-0003-2130-8832}}
\email{pksahoo@hyderabad.bits-pilani.ac.in}
\affiliation{Department of Mathematics, Birla Institute of Technology and
Science-Pilani,\\ Hyderabad Campus, Hyderabad-500078, India.}

%
\date{\today}

\begin{abstract}
This paper examines wormhole geometries in the context of $f(Q, T)$ gravity under the background of non-commutative distributions. We discuss the analytical solutions assuming spherical symmetry and the presence of conformal Killing vectors, which provides a systematic approach for seeking exact wormhole solutions. Specifically, the imposition of conformal symmetry places noteworthy constraints on the model, shaping the analytical outcomes more precisely. We studied the properties of traversable wormholes under both Gaussian and Lorentzian distributions and noticed that NEC and SEC are violated in the neighborhood of the wormhole throat. We also observed the influence of model parameters as well as non-commutative parameters for these violations. Employing the ``volume integral quantifier", it is established that conformally symmetric wormhole geometries may, in principle, be constructed with infinitesimally small amounts of matter, violating the averaged null energy condition. Further, equilibrium forces and the complexity factor of the non-commutative distributed wormholes have also been explored.
\end{abstract}

\maketitle



\section{Introduction}\label{sec1}
\indent Wormholes, frequently featured in science fiction, serve as tunnel-like structures connecting different universes or widely separated regions within the same universe. These geometric constructs are conceptualized as potential means for rapid interstellar travel, warp drives, and time machines. In the mid-1930s, Einstein and Rosen \cite{Einstein1} explored this direction, constructing an elementary particle model represented by a bridge connecting two identical sheets known as the Einstein-Rosen bridge (ERB). However, this mathematical representation of physical space proved unsuccessful as a particle model. Two decades later, Wheeler \cite{Einstein2} delved into topological considerations within general relativity (GR), introducing the concept of ``gravitational-electromagnetic entity" or geons. Geons were predicted as configurations of the gravitational field, potentially coupled to other zero-mass fields like massless neutrinos \cite{Einstein3} or the electromagnetic field \cite{Einstein4}. In 1962, Fuller and Wheeler \cite{Einstein5} demonstrated that the ERB would instantaneously collapse upon formation, confirming it is a non-traversable wormhole, even for photons.\\
\indent Modern attention has been directed toward traversable Lorentzian wormholes, which were initially proposed by Morris and Thorne \cite{Thorne/1988} and subsequently developed by Morris, Thorne, and Yurtsever \cite{Thorne/1988a}. Their foundational work introduced a static, spherically symmetric metric that connects two asymptotically flat space-times, allowing for the unrestricted transit of matter and radiation. This concept has since become a well-established solution within the framework of GR. However, the wormhole solutions exhibit asymptotically flat geometries with a radius that may be either constant or variable, contingent upon their specific configurations. Consequently, these geometries possess a minimum surface area, which is intricately tied to satisfying the flare-out conditions characteristic of the wormhole's throat. To achieve this property, it is postulated that the corresponding space-times necessitate a stress-energy tensor that violates the weak/null energy conditions. Within the framework of classical GR, this violation implies that the matter responsible for the creation of the wormhole must exhibit exceedingly exotic properties, such as negative energy matter \cite{Visser1}, including ghost scalar fields or phantom energy \cite{Armendariz, Lobo1, Sushkov}. While this notion of exotic matter may initially seem peculiar, it finds its roots in the domain of quantum field theory. It emerges as a natural consequence when considering the fluctuations in the topology of space-time over time \cite{Wheeler3}.\\
The exploration of wormhole geometries has gained significant interest, not only within the realm of modified theories of gravity but also in the context of higher-dimensional gravitational theories \cite{Mak, Zangeneh, Galiakhmetov, Kar2, Ziaie}, including the domain of Kaluza-Klein gravity \cite{Singleton, Leon, Folomeev}. These theories offer several advantages, including bypassing nonstandard fluids, which has been a primary motivation for extensive research within modified gravity theories. Furthermore, the specific modifications applied to Einstein's gravity provide additional degrees of freedom within the gravitational sector, presenting opportunities to address challenges associated with dark energy and dark matter. Within the framework of modified gravity with $f(R)$ theories, the cosmic evolution of wormhole geometries has been investigated in \cite{Pavlovic, Idris}. In a related context, Mazharimousavi and Halilsoy \cite{Halilsoy} have comprehensively examined the conditions under which wormholes exist in both vacuum and non-vacuum scenarios. They have successfully derived stable wormhole geometries for the $f(R)$ model, focusing on polynomial evolution. Further, one can read some recent interesting investigations of wormhole geometry in different modified gravity theories \cite{Karakasis, Golchin, Eid, Goswami, Malik, Ahmad, Bhatti, Chanda, Rosa, Tayde 1, Tayde 2, Boehmer, Rani, Momeni}.\\
\indent A recent development in the field of modified gravity is the introduction of matter-geometry coupling through the formulation of the $f(Q, T)$ theory, as presented in \cite{Y.Xu}. This theory allows for any viable function involving the non-metricity scalar Q and the trace of the energy-momentum tensor T to serve as the Lagrangian. While $f(Q, T)$ gravity is a novel approach, it has demonstrated utility in various cosmological contexts. The initial cosmological implications of $f(Q, T)$ gravity were explored in \cite{Y.Xu}, and subsequent studies in \cite{Arora111} delved into the investigation of late-time accelerated expansion, considering observational constraints within the framework of this gravity model. Further research has extensively explored applications in Cosmological inflation \cite{Shiravand}, Baryogenesis \cite{Bhattacharjee}, Cosmological perturbations \cite{Najera}, and the Reconstruction of $f(Q, T)$ Lagrangian \cite{Gadbail}. Notably, there has been limited exploration of the astrophysical scenario within this modified gravity framework. Addressing this gap, \cite{Tayde12} focused on investigating static spherically symmetric wormhole solutions in $f(Q, T)$ gravity, considering both linear and non-linear models under different equations of state relations. Their findings confirmed that while exact solutions could be analytically obtained for the linear model, achieving the same for the non-linear model proved to be a substantially challenging task. Recently, constant-roll and primordial black holes in $f(Q, T)$ gravity have been discussed in \cite{Bourakadi1}. Despite increasing attention towards $f(Q, T)$ gravity, it is a relatively emergent theory with considerable unused potential. Given the limited exploration in the astrophysical domain within this gravity theory, we are motivated to contribute by investigating wormhole solutions supported by non-commutative geometry in the context of $f(Q, T)$ gravity.
\\
\indent Fundamentally, non-commutative geometry is an inherent characteristic of the manifold itself, as elucidated in \cite{Smailagic1}, and it can be incorporated into GR through modifications of the matter source. Synder initially introduced the concept of non-commutative space-time \cite{Snyder1, Snyder2} to address issues related to divergences in relativistic quantum field theory. Schneider and DeBenedictis comprehensively investigated the background of non-commutative distributions in \cite{Smailagic2}. Non-commutative geometry also aspires to establish a unified framework in which GR and the Standard Model can be placed on a similar base, enabling the description of gravity, electro-weak, and strong forces as manifestations of gravitational interactions within a single unified space-time \cite{Stephan}. A significant improvement within string/M-theory has involved the requirement for space-time quantization, wherein space-time coordinates are treated as noncommuting operators on a D-brane \cite{Witten1, Witten2}. This non-commutativity of space-time is represented by the commutator
\begin{equation}\label{1a}
[x_\mu, x_\nu] = i \Theta_{\mu\nu},
\end{equation}
where $\Theta_{\mu\nu}$ is an antisymmetric matrix defining the fundamental discretization of space-time. Additionally, it has been demonstrated that non-commutativity replaces point-like structures with smeared objects within a flat space-time context \cite{Spallucci3}. Consequently, the possibility arises that non-commutativity could offer a potential solution to the divergences encountered in GR. The smearing effect is mathematically implemented by substituting the Dirac delta function with a Gaussian distribution characterized by a minimal length of $\sqrt{\Theta}$ where $\Theta$ has the dimension $length^2$.\\
In recent years, non-commutative geometry has garnered significant attention among researchers, emerging as a pivotal property in space-time geometry with substantial implications across various domains. The contribution of non-commutative geometry to the field of wormholes is that it provides a mathematical framework for studying the properties of compact objects such as wormholes. Parikh-Wilczek Tunneling from Non-commutative Higher Dimensional Black Holes has been investigated in Ref. \cite{Nozari1}. Sushkov explored wormholes supported by phantom energy utilizing Gaussian distribution, as discussed in \cite{Nozari2}. Rahaman et al. \cite{Nozari3} investigated wormhole solutions within a background characterized by a Gaussian distribution, determining their existence solely in four and five dimensions. The BTZ black hole investigation under non-commutative backgrounds was explored in \cite{Nozari4}. Moreover, wormhole solutions have been discussed under the non-commutative background in $f(T)$ \cite{Rani1} as well as $f(Q)$ gravity \cite{Rani2}.
\\
\indent Finding exact solutions to the Einstein field equations can be challenging unless specific symmetry restrictions are imposed on the space-time geometry. These limitations are typically described in terms of isometries, also known as Killing vectors, possessed by the space-time metric. These symmetries, often referred to as collineations and defined by the equation
\begin{equation}\label{1b}
 \mathcal{L}_\xi \psi=\Psi,   
\end{equation}
where $\xi$ represents the collineation (symmetry) vector, $\mathcal{L}$ denotes the Lie derivative. $\psi$ denotes the tensor field and it can be $g_{\mu\nu}$, $R_{\mu\nu}$, $R^\eta_{\mu\nu\sigma}$, $\gamma^{\sigma}_{\mu\nu}$. Also, $\Psi$ is the tensor with the same index symmetries as $\psi$. Among these, conformal Killing vectors (CKVs) offer a particularly insightful approach to understanding space-time geometry, and it can be derived by setting $\psi=g_{\mu\nu}$ and $\Psi=F g_{\mu\nu}$, where $F$ represents conformal factor. This technique presents an inherent symmetry that allows obtaining exact solutions from highly non-linear field equations. A detailed description of conformal symmetry can be found in \cite{Hall}. A notable finding indicates that in the case of a static metric, neither $\xi$ nor $F$ necessarily need to remain static. Researchers explored this observation in \cite{Herrera1}, who employed this finding to demonstrate that the equation of state (EoS) can be uniquely determined by the Einstein equations for a specific set of conformal motions. Subsequently, Maartens and Maharaj \cite{Herrera2} extended upon this particular solution, focusing on the static sphere of charge imperfect fluid. Moreover, in \cite{Herrera3}, Kuhfitting studied the barotropic equation of state that allows a one-parameter group of conformal motion for the wormhole. Recently, wormholes have been investigated with conformal symmetry under the assumption of non-commutative geometries in modified symmetric teleparallel gravity \cite{Singh2}. However, based on our knowledge, there is no work on wormhole geometry based on conformal symmetry in $f(Q, T)$ gravity. This mathematical approach may be beneficial for deducing exact solutions from complex, non-linear partial differential equations. Therefore, examining wormhole solutions using CKVs in the context of $f(Q, T)$ gravity would be interesting.\\
The structure of this article is outlined as follows: In Section \ref{sec2}, we introduce the fundamental formalism of $f(Q, T)$ gravity and derive the corresponding field equations. Additionally, Section \ref{sec3} provides a brief introduction to conformal symmetry. A concise review of non-commutative geometry with the linear form of $f(Q, T)$ is presented in Section \ref{sec4}. To assess the quantity of exotic matter, we employ a volume integral quantifier parameter in Section \ref{sec5} and examine the complexity factor in Section \ref{sec6}. Furthermore, the stability of the obtained wormhole solution is verified using the TOV equation in Section \ref{sec7}. Finally, we summarize our findings in the concluding Section \ref{sec8}.
\section{Basic Criteria of a traversable wormhole and formalism of $f(Q, T)$ gravity}
\label{sec2}
We consider the wormhole metric in the Schwarzschild coordinates $(t,\,r,\,\theta,\,\Phi)$ defined by \cite{Visser1, Thorne/1988}.
\begin{equation}\label{10}
ds^2=e^{2\phi(r)}dt^2-\left(1-\frac{b(r)}{r}\right)^{-1}dr^2-r^2\,d\theta^2-r^2\,\sin^2\theta\,d\Phi^2\,,
\end{equation}
where $b(r)$ stands for the shape function defining the wormholes' shape. The function $\phi(r)$ denotes the redshift function related to the gravitational redshift. Also, in order to make a wormhole to become traversable, the shape function $b(r)$ should fulfill the flaring-out condition, as specified by $(b-b'r)/b^2>0$ \cite{Thorne/1988}. At the wormhole throat $b(r_0)=r_0$, the condition $b^{\,\prime}(r_0)<1$ is imposed (where $r_0$ gives the throat radius). In addition, the asymptotic flatness condition, i.e.,  as $r\rightarrow \infty$, the ratio $\frac{b(r)}{r}$ should tend to $0$, is also required. Besides, to circumvent the event horizon, $\phi(r)$ must remain finite at all points. Meeting the conditions mentioned earlier could ensure the existence of exotic matter at the throat of the wormhole within the framework of Einstein's GR.\\
We will briefly overview the $f(Q, T)$ gravity. We are examining the action for symmetric teleparallel gravity as introduced in \cite{Y.Xu}
\begin{equation}\label{1}
\mathcal{S}=\int\frac{1}{16\pi}\,f(Q, T)\sqrt{-g}\,d^4x+\int \mathcal{L}_m\,\sqrt{-g}\,d^4x\, ,
\end{equation}
where $f(Q,T)$ is a arbitrary function of $Q$ i.e. non-metricity scalar and a trace of the energy-momentum tensor $T$, $\mathcal{L}_m$ is the matter Lagrangian density, and $g$ denotes the determinant of the metric tensor $g_{\mu\nu}$\\
The equation that defines the non-metricity tensor is as follows \cite{Jimenez}
\begin{equation}\label{2}
Q_{\lambda\mu\nu}=\bigtriangledown_{\lambda} g_{\mu\nu}\,.
\end{equation}
Furthermore, the non-metricity conjugate or superpotential can be formally defined as
\begin{equation}\label{3}
P^\alpha\;_{\mu\nu}=\frac{1}{4}\left[-Q^\alpha\;_{\mu\nu}+2Q_{(\mu}\;^\alpha\;_{\nu)}+Q^\alpha g_{\mu\nu}-\tilde{Q}^\alpha g_{\mu\nu}-\delta^\alpha_{(\mu}Q_{\nu)}\right].
\end{equation}
Traces of the non-metricity tensor can be given by
\begin{equation}
\label{4}
\tilde{Q}_\alpha=Q^\mu\;_{\alpha\mu}\,,\;Q_{\alpha}=Q_{\alpha}\;^{\mu}\;_{\mu}.
\end{equation}
The representation of the non-metricity scalar is as follows \cite{Jimenez}
\begin{eqnarray}
\label{5}
Q &=& -P^{\alpha\mu\nu}\,Q_{\alpha\mu\nu}\\
&=& g^{\mu\nu}\left(L^\beta_{\,\,\,\alpha\beta}\,L^\alpha_{\,\,\,\mu\nu}-L^\beta_{\,\,\,\alpha\mu}\,L^\alpha_{\,\,\,\nu\beta}\right),
\end{eqnarray}
The disformation tensor is denoted by $L^\beta_{\,\,\,\mu\nu}$ and given by 
\begin{equation}\label{6}
L^\beta_{\,\,\,\mu\nu}=\frac{1}{2}Q^\beta_{\,\,\,\mu\nu}-Q_{(\mu\,\,\,\,\,\,\nu)}^{\,\,\,\,\,\,\beta}.
\end{equation}

The gravitational equations of motion can be derived through variation of the action concerning the metric tensor $g_{\mu\nu}$, and they are expressed as follows:
\begin{multline}\label{7}
\frac{-2}{\sqrt{-g}}\bigtriangledown_\alpha\left(\sqrt{-g}\,f_Q\,P^\alpha\;_{\mu\nu}\right)-\frac{1}{2}g_{\mu\nu}f + f_T \left(T_{\mu\nu} +\Psi_{\mu\nu}\right) \\
-f_Q\left(P_{\mu\alpha\beta}\,Q_\nu\;^{\alpha\beta}-2\,Q^
{\alpha\beta}\,\,_{\mu}\,P_{\alpha\beta\nu}\right)=8\pi T_{\mu\nu},
\end{multline}

where $f_Q=\frac{\partial f}{\partial Q}$ and $f_T=\frac{\partial f}{\partial T}$.

The representation of the energy-momentum tensor in the context of a fluid description of space-time is given by
\begin{equation}\label{8}
T_{\mu\nu}=-\frac{2}{\sqrt{-g}}\frac{\delta\left(\sqrt{-g}\,\mathcal{L}_m\right)}{\delta g^{\mu\nu}},
\end{equation}
and
\begin{equation}\label{9}
\Psi_{\mu\nu}=g^{\alpha\beta}\frac{\delta T_{\alpha\beta}}{\delta g^{\mu\nu}}.
\end{equation}
Moreover, In this paper, we assume the diagonal energy-momentum tensor for an anisotropic fluid, which can be read as
\begin{equation}\label{11}
T_{\mu}^{\nu}=\text{diag}[\rho,-p_r,-p_t,-p_t],
\end{equation}
where $\rho$, $p_r$, and $p_t$ represent the energy density, radial pressure, and tangential pressure, respectively.
Further, we adopt the matter Lagrangian $\mathcal{L}_m=-P$ as proposed in \cite{Correa}, resulting in the form for Eq. \eqref{9} as:
\begin{equation}\label{12}
\Psi_{\mu\nu}=-g_{\mu\nu}\,P-2\,T_{\mu\nu}\,,
\end{equation}
where $P$ denotes the total pressure and can be expressed as $P=\frac{p_r+2 p_t}{3}$.
The non-metricity scalar $Q$ for the metric \eqref{10} is defined as given in \cite{Tayde12}
\begin{equation}\label{13}
Q=-\frac{b}{r^2}\left[2\phi^{'}+\frac{rb^{'}-b}{r(r-b)}\right].
\end{equation}\\
The field equations governing $f(Q, T)$ gravity are outlined in the following form \cite{Tayde12}
\begin{multline}\label{14}
8 \pi  \rho =\frac{(r-b)}{2 r^3} \left[f_Q \left(\frac{(2 r-b) \left(r b'-b\right)}{(r-b)^2}+\frac{b \left(2 r \phi '+2\right)}{r-b}\right)
\right. \\ \left.
+\frac{f r^3}{r-b}-\frac{2r^3 f_T (P+\rho )}{(r-b)}+\frac{2 b r f_{\text{QQ}} Q'}{r-b}\right],
\end{multline}
\begin{multline}\label{15}
8 \pi  p_r=\frac{(r-b)}{2 r^3} \left[-f_Q \left(\frac{b }{r-b}\left(\frac{r b'-b}{r-b}+2+2 r \phi '\right)
\right.\right. \\ \left.\left.
-4 r \phi '\right)-\frac{f r^3}{r-b}+\frac{2r^3 f_T \left(P-p_r\right)}{(r-b)}-\frac{2 b r f_{\text{QQ}} Q'}{r-b}\right],
\end{multline}
\begin{multline}\label{16}
8 \pi  p_t=-\frac{(r-b)}{4 r^2} \left[f_Q \left(\frac{\left(r b'-b\right) \left(\frac{2 r}{r-b}+2 r \phi '\right)}{r (r-b)}+
\right.\right. \\ \left.\left.
\frac{4 (2 b-r) \phi '}{r-b}-4 r \left(\phi '\right)^2-4 r \phi ''\right)+\frac{2 f r^2}{r-b}
\right.\\\left.
-4 r f_{\text{QQ}} Q' \phi '-\frac{4r^2 f_T \left(P-p_t\right)}{(r-b)}\right].
\end{multline}
Applying these precise field equations makes it possible to investigate various wormhole solutions within the domain of $f(Q, T)$ gravity models.

\subsection{Energy conditions}
The classical energy conditions, derived from the Raychaudhuri equations, serve as a basis for discussing physically realistic matter configurations. The four commonly utilized energy conditions are weak, null, strong, and dominant. The null energy condition holds particular significance in wormhole solutions in GR. This stems from its direct connection to the energy density necessary to maintain the openness of the wormhole throat. A violation of the null energy condition near the wormhole's throat implies the existence of exotic matter with negative energy density, which is not a characteristic of standard matter sources. Energy conditions impose restrictions on the stress-energy tensor, describing the distribution of matter and energy in space-time, and can be expressed as follows:\\
$\bullet$ The weak energy condition (\textbf{WEC}) :
$\rho\geq0$,\,\, $\rho+p_t\geq0$,\,\, and \,\, $\rho+p_r\geq0$.\\
$\bullet$ The null energy condition (\textbf{NEC}) : $\rho+p_t\geq0$\,\, and \,\, $\rho+p_r\geq0$.\\
$\bullet$ The dominant energy condition (\textbf{\textbf{DEC}}) :  $\rho\geq0$,\,\, $\rho+p_t\geq0$,\,\, $\rho+p_r\geq0$,\,\, $\rho-p_t\geq0$,\,\, and \,\, $\rho-p_r\geq0$.\\
$\bullet$ The strong energy condition (\textbf{SEC}) :
 $\rho+p_t\geq0$,\,\, $\rho+p_r\geq0$,\,\, and \,\, $\rho+p_r+2p_t\geq0$.\\
To conclude, energy conditions offer valuable limitations on the behavior of matter in the cosmos and hold a pivotal role in exploring wormholes.

\section{Conformal symmetry}\label{sec3}
While numerical computations have achieved considerable success, the significance of exact solutions persists in both GR and modified gravity. Exact solutions offer global acceptance without specifying particular parameters and initial conditions. Moreover, conformal symmetries are crucial in providing valuable insights and information regarding the general properties of self-gravitating matter configurations. Conformal symmetry is a symmetry of spacetime that preserves angles but not necessarily distances. The motivation behind using conformal symmetry is that it simplifies the field equations and makes them more tractable. By imposing conformal symmetry on the metric, the number of independent variables in the field equations is reduced, which makes it easier to find exact solutions. Motivated by these considerations, we assume that the spherically symmetric and static space-time allows for a conformal motion. This assumption, as detailed in Refs. \cite{Herrera1, Herrera2} will enable us to simplify the problem and establish its fundamental mathematical structure.\\
In general, conformal motion (CM) is a map  $M\rightarrow M$ such that the metric $g$ of the space-time transforms under the rule
 \begin{equation}
g \rightarrow \tilde{g}=2 e^{F} g, \quad \quad \quad \text{with} \quad \quad F=F(x^a),
 \end{equation}
which can be represented as Eq. \eqref{1b}. In the study presented in \cite{Herrera1}, the authors assumed that the vector field responsible for generating conformal symmetry is spherically symmetric and static in the context of GR. This assumption leads to the expression
 \begin{equation}
\xi=\xi^0 r \frac{\partial}{\partial t}+\xi^1 r \frac{\partial}{\partial r}.
 \end{equation}
 By employing this specific form of the conformal vector in Eq. \eqref{1b}, the derived results include
\begin{eqnarray*}
\xi^{1}\nu^{'}(r)=F(r),\;\;\;\;\;\;
\xi^{1}=\frac{r F(r)}{2},\;\;\;\;\;\;
\xi^{1}\lambda^{'}(r)+2\xi^{1}_{,1}=F(r).
\end{eqnarray*}
Through the resolution of the system, as mentioned above, using the space-time Equation (\ref{10}), we derive the subsequent relationships:
\begin{equation}
e^{2\phi(r)}=K_{1}^{2}r^2, \;\;\;\;e^{\lambda (r)}=\left(\frac{r-b(r)}{r}\right)^{-1}=\frac{K^{2} _{2}}{(F(r))^{2}}.\label{16a}
\end{equation}
Here, $K_1$ and $K_2$ are regarded as constants of integration.
\section{Wormhole solutions using linear $f(Q,T)$ model}
\label{sec4}
In this section, we will examine a specific and intriguing $f(Q, T)$ model defined as:
\begin{equation}\label{17}
f(Q,T)=\alpha Q+\beta T\,.
\end{equation}
Here, $\alpha$ and $\beta$ represent model parameters. This model was initially introduced by Xu et al. \cite{Y.Xu} and naturally characterizes an exponentially expanding Universe, with $\rho \propto e^{-H_0 t}$ \cite{Y.Xu}. It has also been subject to constraints from observational data regarding the Hubble parameter, as discussed in \cite{Arora11}. Additionally, Loo et al. \cite{Loo1} employed this model to investigate Bianchi type-I cosmology using observational datasets, including Type Ia supernovae and Hubble. Furthermore, \cite{Tayde12} has investigated wormhole solutions for this model, considering different equations of state (EoS) relations. In this study, we aim to assess the viability of this model within the context of non-commutative geometries. When applying the linear form \eqref{17} with a non-constant redshift function, the corresponding field equations \eqref{14}-\eqref{16} can be represented as:
\begin{multline}\label{18}
\rho =\frac{\alpha}{\mathcalboondox{L}_1} \left(-(\beta +48 \pi ) r F F'-8 (\beta +3 \pi ) F^2+(12 \pi -\beta )
 \right. \\ \left.
 \times 2 {K_2}^2\right)\,,
 \end{multline}
 \begin{equation}\label{19}
 p_r=\frac{\alpha}{\mathcalboondox{L}_1} \left(13 \beta  r F F'+(72 \pi -4 \beta ) F^2+2 (\beta -12 \pi ) {K_2}^2\right)\,,
 \end{equation}
 \begin{equation}\label{20}
 p_t=\frac{\alpha}{\mathcalboondox{L}_1} \left((\beta +48 \pi ) r F F'+8 (\beta +3 \pi ) F^2-4 \beta  {K_2}^2\right)\,,
 \end{equation}
 where, $\mathcalboondox{L}_1=6 (\beta +8 \pi )  (4 \pi -\beta ) {K_2}^2 r^2$.\\
In the upcoming subsections, we will analyze wormhole solutions within the context of Gaussian and Lorentzian distributions. Our objective is to examine the influence of these non-commutative geometries on the wormhole solutions by employing a conformal approach through energy conditions and the behaviors of shape functions.
 
\subsection{Gaussian distribution}
\label{subsubsec1}
\begin{figure*}
\centering
\includegraphics[width=14.5cm,height=6cm]{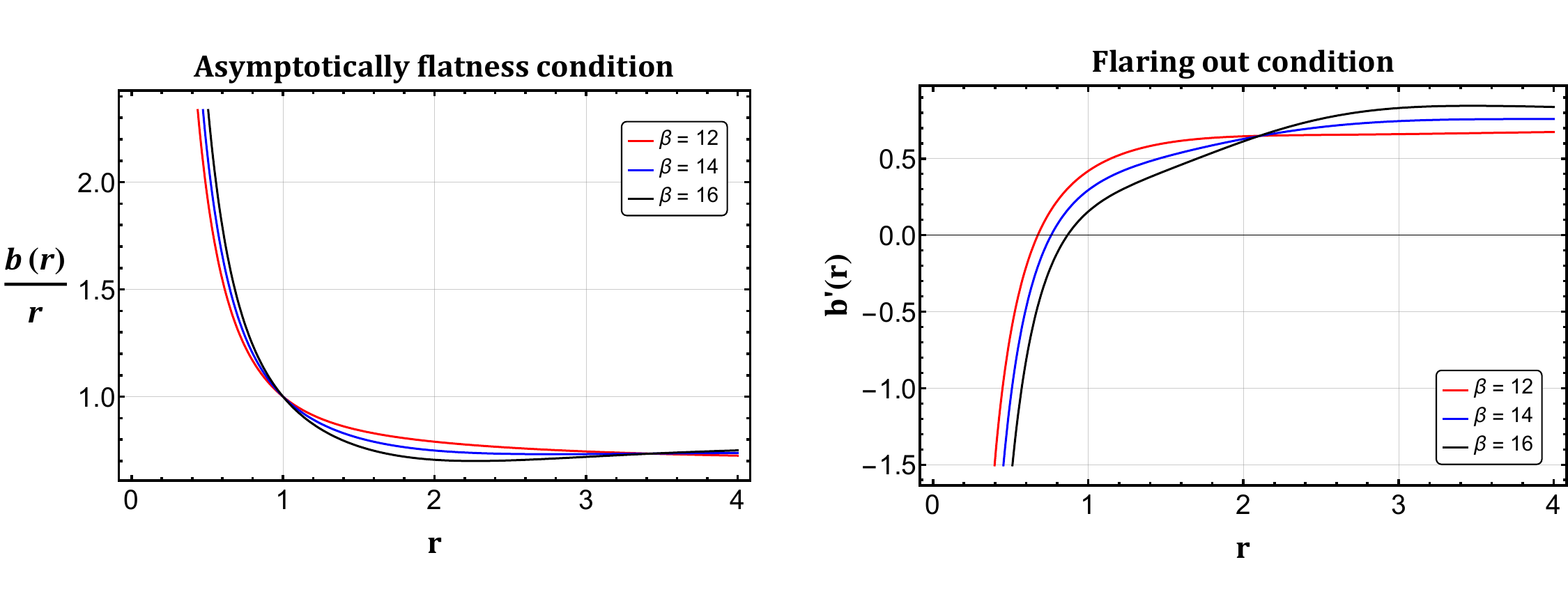}
\caption{The figure displays the variations in the asymptotically flatness condition \textit{(on the left)} and the flare-out condition \textit{(on the right)} as a function of the radial coordinate `$r$ ' for various values of `$\beta$ '. Additionally, we maintain fixed values for other parameters, including $\alpha=0.3,\, \Theta=0.5,\, M=0.25,\, K_2=0.2,\, \text{and} \, r_0 = 1$.}
\label{fig1}
\end{figure*}

\begin{figure*}
\centering
\includegraphics[width=14.5cm,height=6cm]{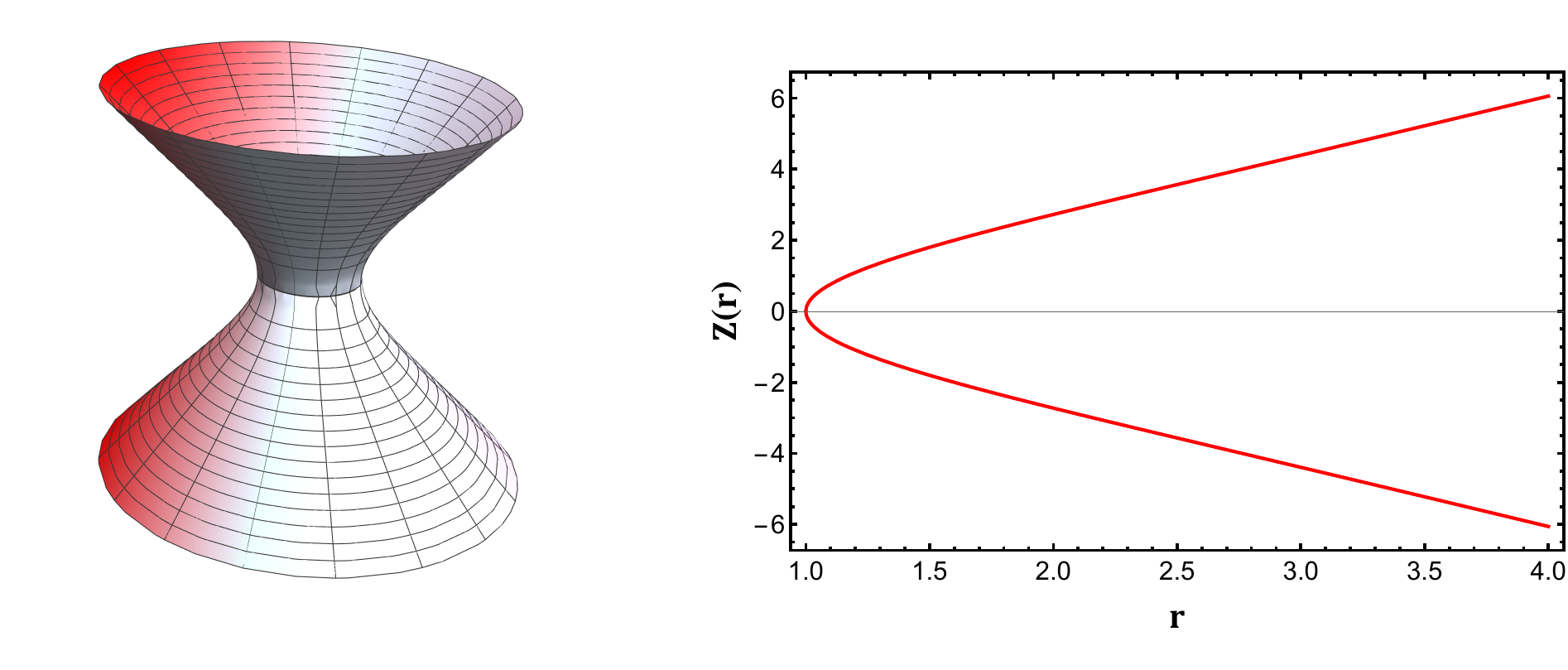}
\caption{The figure displays the embedding diagram for Gaussian distribution. Additionally, we maintain fixed values for other parameters, including $\alpha=0.3,\, \Theta=0.5,\, M=0.25,\, K_2=0.2,\,\beta=14,\, \text{and} \, r_0 = 1$.}
\label{fig15}
\end{figure*}

This subsection focuses on the mass density of a static, spherically symmetric, smeared, particle-like gravitational source, described as \cite{P. Nicoloni, A. Smailagic1}
 \begin{equation}\label{21}
 \rho =\frac{M e^{-\frac{r^2}{4 \Theta }}}{8 \pi ^{3/2} \Theta ^{3/2}}\,.
 \end{equation}
Instead of being strictly localized at a single point, the particle mass $M$ is spread over an area with a direct measure of $\sqrt{\Theta}$. This arises due to the uncertainty encoded within the coordinate commutator.\\
 By contrasting Equations \eqref{18} and \eqref{21}, we can establish the differential equation for the Gaussian distribution as follows
 \begin{multline}\label{23}
 \frac{\alpha}{\mathcalboondox{L}_1} \left(-(\beta +48 \pi ) r F F'-8 (\beta +3 \pi ) F^2
 \right. \\ \left.
 +2 (12 \pi -\beta ){K_2}^2\right)=\frac{M e^{-\frac{r^2}{4 \Theta }}}{8 \pi ^{3/2} \Theta ^{3/2}}\,.
\end{multline}
By solving the aforementioned expressions, we acquire the following result
\begin{multline}\label{23a}
F(r)=\left(r^{-2\mathcalboondox{L}_3} \left({K_2}^2 \mathcalboondox{L}_2 r^{2(\mathcalboondox{L}_3 +1)} E_{-\mathcalboondox{L}_3}\left(\frac{r^2}{4 \Theta }\right)+ c_1
 \right. \right. \\ \left.\left.
+\frac{(12 \pi -\beta ) {K_2}^2 }{4 (\beta +3 \pi )}r^{2\mathcalboondox{L}_3}\right)\right)^{1/2}\,,
\end{multline}
 where $\mathcalboondox{L}_2=\frac{3 (4 \pi -\beta ) (\beta +8 \pi ) M}{4 \pi ^{3/2} \alpha  (\beta +48 \pi ) \Theta ^{3/2}}$, $\mathcalboondox{L}_3=\frac{8 (\beta +3 \pi )}{\beta +48 \pi }$ and $c_1$ is the integrating constant. ``E" is an exponential integral function and defined by $E_n(z)=\int _1^{\infty }\frac{e^{-zt}}{t^n}dt $.
Therefore, the shape function can be derived using Eqs. \eqref{16a} and \eqref{23a} as 
 \begin{multline}\label{a1}
b(r)=r \left(\frac{5 \beta }{4 (\beta +3 \pi )}-\mathcalboondox{L}_2 r^2 E_{-\mathcalboondox{L}_3}\left(\frac{r^2}{4 \Theta }\right)-\frac{c_1 r^{-2\mathcalboondox{L}_3}}{{K_2}^2}\right)\,.
 \end{multline}
To determine the value of $c_1$, we apply the throat condition $b(r_0)=r_0$ to the previously mentioned expression \eqref{a1}, resulting in
\begin{equation}\label{a2}
c_1= {r_0}^{\frac{15 \beta }{\beta +48 \pi }} {K_2}^2\left(-\mathcalboondox{L}_2 {r_0}^3 E_{-\mathcalboondox{L}_3}\left(\frac{{r_0}^2}{4 \Theta }\right)+\frac{5 r_0 \beta }{4 (\beta +3 \pi )}-r_0\right)\,.
 \end{equation}
Consequently, the shape function described in \eqref{a1} can be expressed as 
 \begin{multline}\label{a3}
b(r)=r \left(\frac{5 \beta }{4 (\beta +3 \pi )}-\mathcalboondox{L}_2 r^2 E_{-\mathcalboondox{L}_3}\left(\frac{r^2}{4 \Theta }\right)-r^{-2\mathcalboondox{L}_3} 
\right. \\ \left.
\hspace{1cm}  \times\left({K_2}^2 {r_0}^{\frac{15 \beta }{\beta +48 \pi }} \left(-\mathcalboondox{L}_2 {r_0}^3 E_{-\mathcalboondox{L}_3}\left(\frac{{r_0}^2}{4 \Theta }\right) -{r_0}
\right.\right.\right. \\ \left.\left.\left.
+\frac{5 {r_0} \beta }{4 (\beta +3 \pi )}\right)\right)\right)\,.
 \end{multline}
We will now explain the graphical representation of the derived shape function and the necessary conditions for the existence of a wormhole. We will make appropriate selections for the relevant free parameters to accomplish this. First, we will examine how the conditions of the shape function behave for the Gaussian distribution when the redshift function remains non-constant. The behavior of the asymptotic flatness condition and the fulfillment of the flaring out condition for different values of $\beta$ are visually presented in Fig. \ref{fig1}. One can observe that the left graph in Fig. \ref{fig1} portrays the asymptotic behavior of the shape function for varying values of $\beta$. It demonstrates that as the radial distance increases, the ratio $\frac{b(r)}{r}$ approaches zero, thereby confirming the asymptotic behavior of the shape function. The accompanying right graph shows the satisfaction of the flaring out condition ``$b'(r_0) < 1$" at the wormhole throat. In this instance, we consider the wormhole throat at $r_0=1$.\\
Following the formulation by Morris and Thorne \cite{Thorne/1988}, the embedding surface of the wormhole is expressed by the function $z(r)$, governed by the differential equation:
\begin{equation}\label{6d1}
\frac{dz}{dr}=\pm \frac{1}{\sqrt{\frac{r}{b(r)}-1}}.
\end{equation}
It is notable from the above equation that $\frac{dz}{dr}$ diverges at the throat of the wormhole, leading to the inference that the embedding surface takes a vertical exposure at the wormhole throat. Furthermore, the above equation \eqref{6d1} establishes the relationship 
\begin{equation}\label{6d2}
z(r)=\pm \int_{r_0}^{r} \frac{dr}{\sqrt{\frac{r}{b(r)}-1}}.
\end{equation}
Note that the above integral cannot be accomplished analytically. We adopt the numerical technique to produce the wormhole shape in that case. Now, for the Gaussian distributed wormhole shape function \eqref{a3}, we have presented the embedding diagram $z(r)$ and its complete visualization in Fig. \ref{fig15}.

\begin{figure*}
\centering
\includegraphics[width=14cm,height=5.5cm]{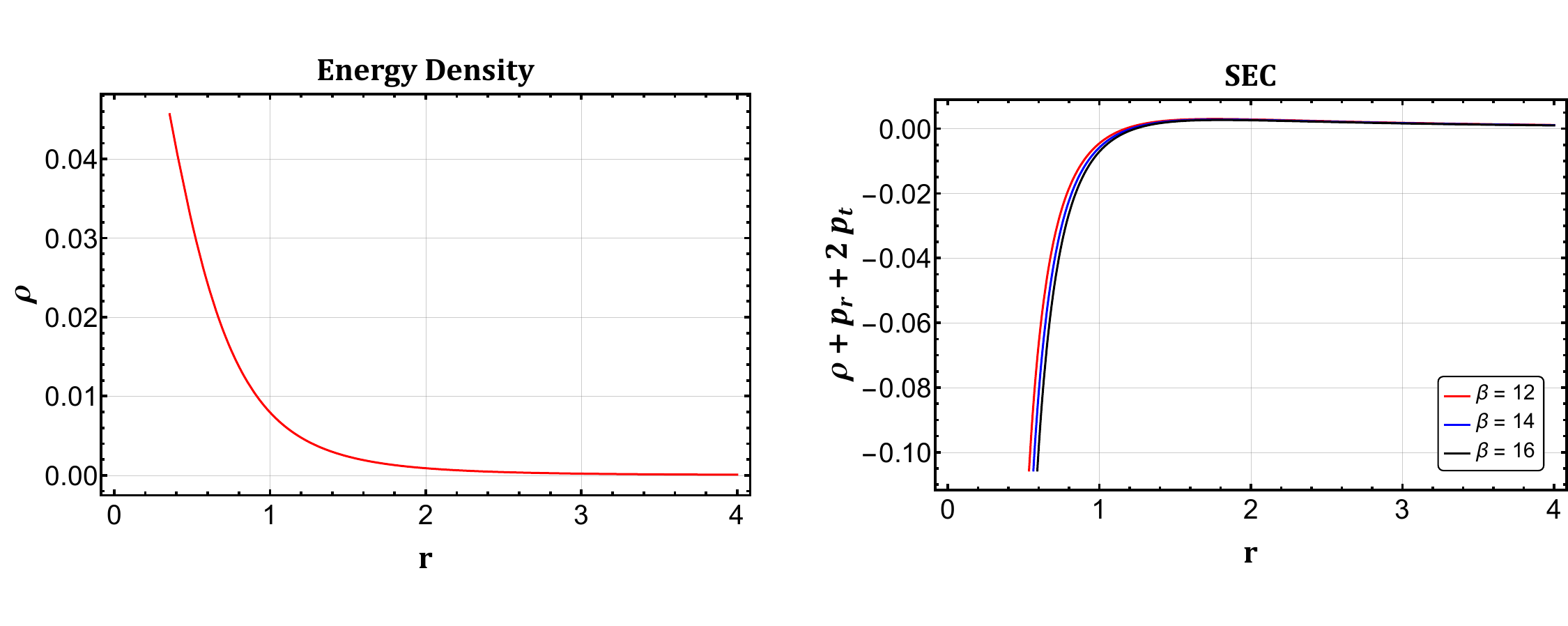}
\caption{The figure displays the variations in the energy density \textit{(on the left)} and the SEC \textit{(on the right)} as a function of the radial coordinate `$r$ ' for various values of `$\beta$ '. Additionally, we maintain fixed values for other parameters, including $\alpha=0.3,\, \Theta=0.5,\, M=0.25,\, K_2=0.2,\, \text{and} \, r_0 = 1$.}
\label{fig2}
\end{figure*}

\begin{figure*}
\centering
\includegraphics[width=14.5cm,height=5cm]{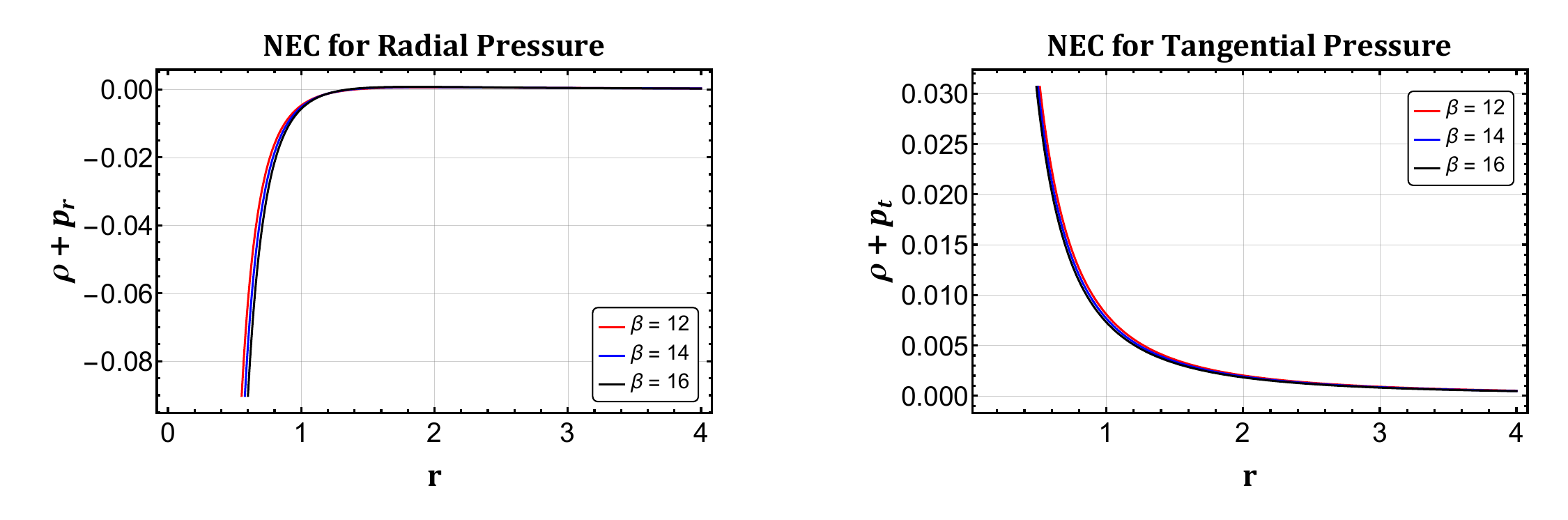}
\caption{The figure displays the variations in the NEC for radial \textit{(on the left)} and tangential \textit{(on the right)} pressure as a function of the radial coordinate `$r$ ' for various values of `$\beta$ '. Additionally, we maintain fixed values for other parameters, including $\alpha=0.3,\, \Theta=0.5,\, M=0.25,\, K_2=0.2,\, \text{and} \, r_0 = 1$.}
\label{fig3}
\end{figure*}

We replace the derived shape function from \eqref{a3} into Eqs to examine the energy conditions, particularly the NEC and SEC. \eqref{19} and \eqref{20}. This allows us to calculate the pressure components as follows:
\begin{multline}\label{a4}
p_r = \frac{r^{-2\mathcalboondox{L}_4}}{\mathcalboondox{M}_1}\left(-1296 (\beta +3 \pi ) \left(\beta ^2+4 \pi  \beta -32 \pi ^2\right)^2 
\right. \\ \left.
\times \Theta  M {r_0}^{2(\mathcalboondox{L}_3 +1)} E_{-\mathcalboondox{L}_3}\left(\frac{{r_0}^2}{4 \Theta }\right)+3 (4 \pi -\beta ) (\beta +48 \pi ) 
\right. \\ \left.
\times \left(-16 \pi ^{3/2} \alpha  (12 \pi -\beta ) \Theta ^{5/2} \left(9 (\beta +8 \pi ) {r_0}^{2\mathcalboondox{L}_3}-r^{2\mathcalboondox{L}_3}
\right.\right.\right. \\ \left.\left.\left.
\times (\beta +48 \pi ) \right)-(\beta +3 \pi ) (\beta +8 \pi ) M r^{2(\mathcalboondox{L}_3 +1)}\left(13 \beta  r^2 
\right.\right.\right. \\ \left.\left.\left.
\times E_{-\mathcalboondox{L}_4}\left(\frac{r^2}{4 \Theta }\right)-36 (\beta +8 \pi ) \Theta  E_{-\mathcalboondox{L}_3}\left(\frac{r^2}{4 \Theta }\right)\right)\right)\right)\,,
\end{multline}
\begin{multline}\label{a5}
p_t = \frac{1}{\mathcalboondox{M}_2}\left(36 (\beta +8 \pi )^2 \Theta  M r^2 E_{-\mathcalboondox{L}_3}\left(\frac{r^2}{4 \Theta }\right)+(\beta +48 \pi ) 
\right. \\ \left.
\times \left(32 \pi ^{3/2} \alpha  \Theta ^{5/2}-(\beta +8 \pi ) M r^4 E_{-\mathcalboondox{L}_4}\left(\frac{r^2}{4 \Theta }\right)\right)\right)\,,
\end{multline}
where  $\mathcalboondox{M}_1=96 \pi ^{3/2} (4 \pi-\beta)(\beta +3 \pi )(\beta +8 \pi)(\beta +48 \pi )^2 \Theta ^{5/2}$, $\mathcalboondox{M}_2=32 \pi ^{3/2} (\beta +8 \pi ) (\beta +48 \pi ) \Theta ^{5/2} r^2$, and $\mathcalboondox{L}_4=\frac{9 (\beta +8 \pi )}{\beta +48 \pi }$.\\
Also, NEC at the throat $r=r_0$ for radial and tangential pressure and SEC are given by
\begin{multline}\label{3210}
\rho + p_r \bigg\vert_{r=r_0}= \frac{1}{\mathcalboondox{L}_5}\left(4 (\beta +48 \pi ) \Theta  \left(\frac{32 \pi ^{3/2} \alpha }{{r_0}^2}(\beta -12 \pi ) 
\right.\right. \\ \left.\left.
\hspace{1.4cm}\times \Theta ^{3/2}+(\beta +8 \pi ) (\beta +48 \pi ) M e^{-\frac{{r_0}^2}{4 \Theta }}\right)+13 \beta  
\right. \\ \left.
\hspace{1cm}\times(\beta +8 \pi ) M \left(36 (\beta +8 \pi ) \Theta  E_{-\mathcalboondox{L}_3}\left(\frac{{r_0}^2}{4 \Theta }\right)
\right.\right. \\ \left.\left.
-{r_0}^2 (\beta +48 \pi ) E_{-\mathcalboondox{L}_4}\left(\frac{{r_0}^2}{4 \Theta }\right)\right)\right)  \,,  
\end{multline}
\begin{multline}
\rho + p_t \bigg\vert_{r=r_0}=\frac{1}{32 \pi ^{3/2} \Theta ^{5/2}}\left(\frac{32 \pi ^{3/2} \alpha  \Theta ^{5/2}}{{r_0}^2 (\beta +8 \pi )}+4 \Theta  M 
\right. \\ \left.
\hspace{0.5cm}\times e^{-\frac{{r_0}^2}{4 \Theta }}+\frac{36 (\beta +8 \pi ) \Theta  M}{\beta +48 \pi}E_{-\mathcalboondox{L}_3}\left(\frac{{r_0}^2}{4 \Theta }\right)
\right. \\ \left.
-{r_0}^2 M E_{-\mathcalboondox{L}_4}\left(\frac{{r_0}^2}{4 \Theta }\right)\right)\,,
\end{multline}
\begin{multline}
\rho + p_r +2 p_t \bigg\vert_{r=r_0}=  \frac{(\beta +8 \pi )}{\mathcalboondox{L}_5}\left(4 (\beta +48 \pi ) \Theta  \left(\frac{48 \pi ^{3/2} }{{r_0}^2}
\right.\right. \\ \left.\left.
\hspace{1.3cm} \times \alpha \Theta ^{3/2}+(\beta +48 \pi ) M e^{-\frac{{r_0}^2}{4 \Theta }}\right) +3 (5 \beta +32 \pi ) M
\right. \\ \left.
\hspace{1.3cm} \times \left(36 (\beta +8 \pi ) \Theta  E_{-\mathcalboondox{L}_3}\left(\frac{{r_0}^2}{4 \Theta }\right)-{r_0}^2 (\beta +48 \pi ) 
\right.\right. \\ \left.\left.
E_{-\mathcalboondox{L}_4}\left(\frac{{r_0}^2}{4 \Theta }\right)\right)\right)\,,
\end{multline}
where $\mathcalboondox{L}_5=32 \pi ^{3/2} (\beta +8 \pi ) (\beta +48 \pi )^2 \Theta ^{5/2}$.\\
Upon analyzing the above expressions, it becomes evident that $\beta$ cannot equal either $-8\pi$ or $-48\pi$. With this in mind, we created energy condition graphs, as illustrated in Figs. \ref{fig2} and \ref{fig3}. The left graph of Fig. \ref{fig2} showcases the energy density as a function of radius ($r$), revealing a consistently decreasing behavior throughout space-time. Conversely, the right graph in the same figure demonstrates the behavior of the SEC for various values of $\beta$, which consistently exhibits a negative trend. Moving on to Fig. \ref{fig3}, the left graph signifies the negative behavior of the radial NEC, indicating a violation of the NEC. In addition, the right graph in Fig. \ref{fig3} affirms the favorable behavior of tangential NEC, which diminishes as we move further from the wormhole throat. This analysis suggests that exotic matter may be preserving wormhole solutions in scenarios involving non-commutative geometry under conformal symmetry with non-metricity-based gravity, which shows the case for the framework of GR.

\subsection{Lorentzian distribution}\label{subsec2}
This subsection will delve into the wormhole solution within non-commutative geometry, specifically utilizing the Lorentzian distribution. The corresponding energy density in this context can be expressed as \cite{P. Nicoloni,Mehdipour11}
 \begin{equation}\label{24}
 \rho =\frac{\sqrt{\Theta } M}{\pi ^2 \left(\Theta +r^2\right)^2}\,.
 \end{equation}
 In the provided equation, $M$ represents the total mass, while $\Theta$ serves as the non-commutativity parameter.\\
Now, through a comparison of Eqs. \eqref{18} and \eqref{24}, we can derive the differential equation associated with the Lorentzian distribution as
\begin{multline}\label{25}
\frac{\alpha}{\mathcalboondox{L}_1} \left(-(\beta +48 \pi ) r F F'-8 (\beta +3 \pi ) F^2
 \right. \\ \left.
 +2 (12 \pi -\beta ){K_2}^2\right)=\frac{\sqrt{\Theta } M}{\pi ^2 \left(\Theta +r^2\right)^2}\,.
 \end{multline}
Upon integration of the aforementioned equation \eqref{23}, the resulting expression is
 \begin{multline}\label{4b1}
F(r)=\left(r^{-2\mathcalboondox{L}_3} \left(\frac{(12 \pi -\beta ) {K_2}^2 }{4 (\beta +3 \pi )}r^{2\mathcalboondox{L}_3}+c_2-{K_2}^2 \mathcalboondox{L}_2 \Theta
\right.\right. \\ \left.\left.
\times \frac{8 r^{2\mathcalboondox{L}_3} }{\mathcalboondox{L}_3\sqrt{\pi } }\left(\, _2F_1\left(1,\mathcalboondox{L}_3,\mathcalboondox{L}_4,-\frac{r^2}{\Theta }\right)
\right.\right.\right. \\ \left.\left.\left.
-\, _2F_1\left(2,\mathcalboondox{L}_3,\mathcalboondox{L}_4,-\frac{r^2}{\Theta }\right)\right)\right)\right)^{1/2}\,,
 \end{multline}
where $c_2$ represents the constant of integration and $_2F_1$ is the hypergeometric function and it can be defined by $_2F_1(a,b,c,z)=\mathlarger{\sum}\limits_{k=0}^{\infty } \frac{a_k b_k z^k}{k! c_k}$. Thus, the shape function can be obtained using Eqs. \eqref{16a} and \eqref{4b1} as
\begin{multline}\label{4b2}
b(r)=-\frac{c_2 r^{-\frac{15 \beta }{\beta +48 \pi }}} K_2^2+\frac{(\beta -12 \pi ) r}{4 (\beta +3 \pi )}+r +\frac{8\mathcalboondox{L}_2 \Theta  r }{\sqrt{\pi }}\\
\times\left(\, _2F_1\left(1,\mathcalboondox{L}_3,\mathcalboondox{L}_4,-\frac{r^2}{\Theta }\right)-\frac{\Theta }{\Theta +r^2}\right)\,.
\end{multline}
To find the value of $c_1$, we use the throat condition $b(r_0)=r_0$ to the above equation, resulting in
\begin{multline}
c_2={r_0}^{\frac{15 \beta }{\beta +48 \pi }}{K_2}^2\left(\frac{r_0 (\beta -12 \pi )}{4 (\beta +3 \pi )}+\frac{8\mathcalboondox{L}_2 \Theta  {r_0} }{\sqrt{\pi }}
\right. \\ \left.
\times\left(\, _2F_1\left(1,\mathcalboondox{L}_3,\mathcalboondox{L}_4,-\frac{{r_0}^2}{\Theta }\right)-\frac{\Theta }{\Theta +{r_0}^2}\right)\right).
\end{multline}
Substituting the calculated value of $c_2$ in Eq. \eqref{4b2}, we get
 \begin{multline}\label{34}
b(r)=-r^{-\frac{15 \beta }{\beta +48 \pi }}{r_0}^{\frac{15 \beta }{\beta +48 \pi }}\left(\frac{r_0 (\beta -12 \pi )}{4 (\beta +3 \pi )}+\frac{8\mathcalboondox{L}_2 \Theta  {r_0} }{\sqrt{\pi }}
\right. \\ \left.
\times\left(\, _2F_1\left(1,\mathcalboondox{L}_3,\mathcalboondox{L}_4,-\frac{{r_0}^2}{\Theta }\right)-\frac{\Theta }{\Theta +{r_0}^2}\right)\right)+\frac{(\beta -12 \pi ) r}{4 (\beta +3 \pi )}\\
+r +\frac{8\mathcalboondox{L}_2 \Theta  r }{\sqrt{\pi }}\left(\, _2F_1\left(1,\mathcalboondox{L}_3,\mathcalboondox{L}_4,-\frac{r^2}{\Theta }\right)-\frac{\Theta }{\Theta +r^2}\right)\,.
 \end{multline}
 \begin{figure*}
\centering
\includegraphics[width=14.5cm,height=6cm]{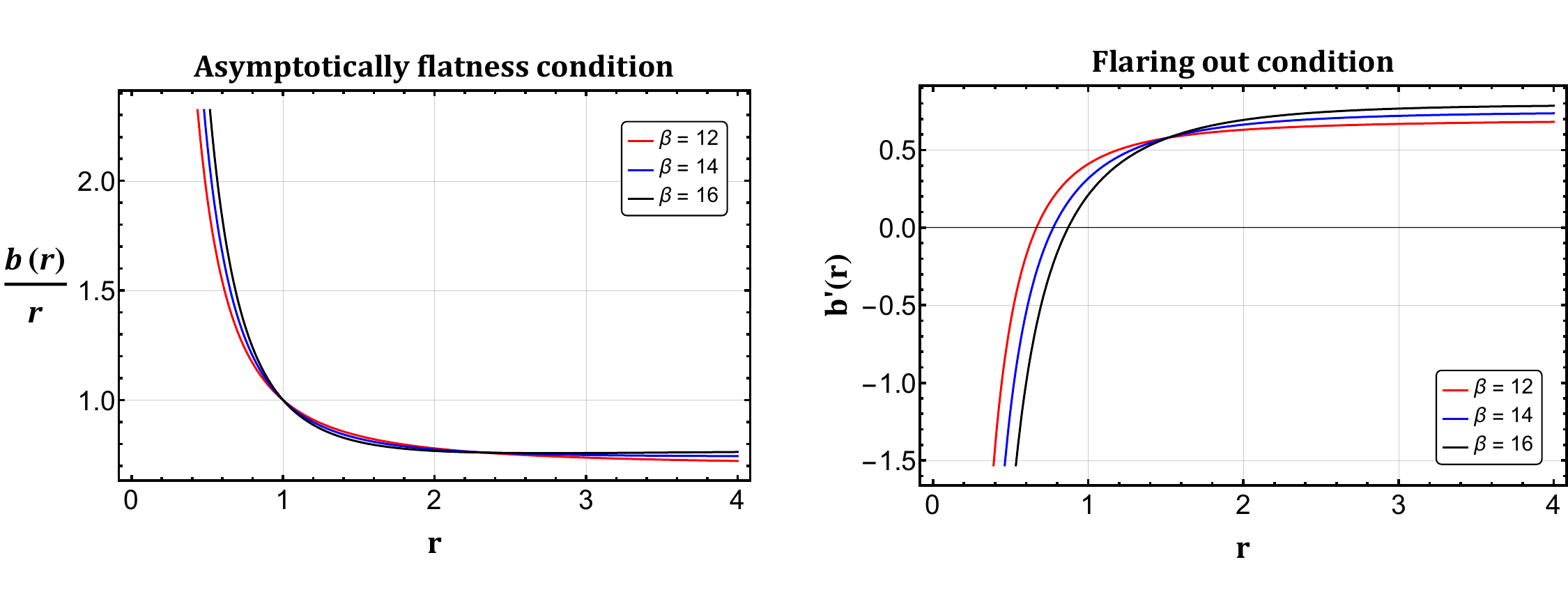}
\caption{The figure displays the variations in the asymptotically flatness condition \textit{(on the left)} and the flare-out condition \textit{(on the right)} as a function of the radial coordinate `$r$ ' for various values of `$\beta$ '. Additionally, we maintain fixed values for other parameters, including $\alpha=0.3,\, \Theta=0.5,\, M=0.25,\, K_2=0.2,\, \text{and} \, r_0 = 1$.}
\label{fig4}
\end{figure*}
 \begin{figure*}
\centering
\includegraphics[width=14.5cm,height=6cm]{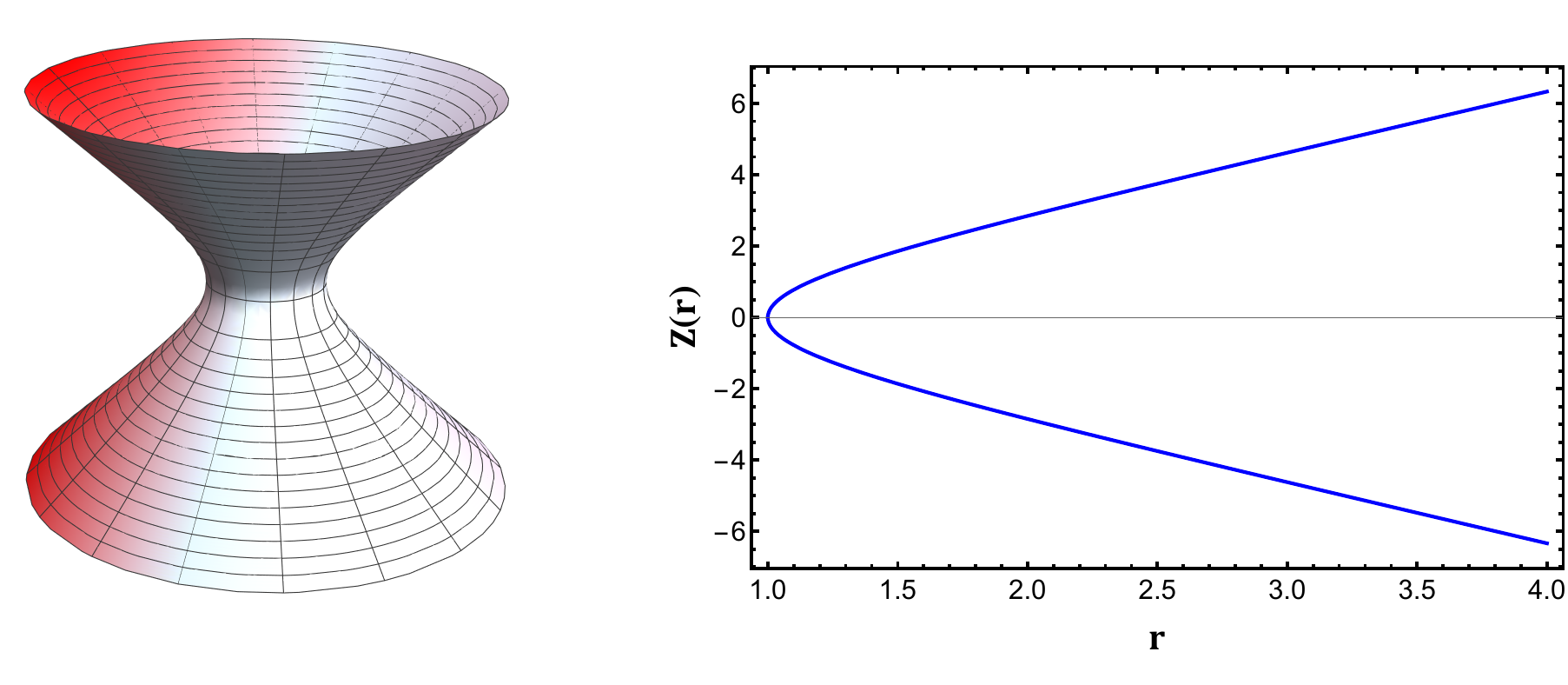}
\caption{The figure displays the embedding diagram for Lorentzian distribution. Additionally, we maintain fixed values for other parameters, including $\alpha=0.3,\, \Theta=0.5,\, M=0.25,\, K_2=0.2,\,\beta=14,\, \text{and} \, r_0 = 1$.}
\label{fig16}
\end{figure*}
We will now delve into the visual representation of the shape function and explore the essential conditions for the existence of a wormhole. To do so, we will carefully select the appropriate parameters for this study. We present the graphical depiction of the asymptotic behavior of the shape function and the fulfillment of the flaring out condition for various values of $\beta$ in Fig. \ref{fig4}. The left graph in Fig. \ref{fig4} provides insight into the asymptotic nature of the shape function for various values of $\beta$. As the radial distance increases, the ratio $\frac{b(r)}{r}$ approaches $0$, affirming the asymptotic nature of the shape function. Further, the corresponding right graph clearly illustrates the satisfaction of the flaring out condition, where $b'(r_0) < 1$, at the wormhole throat. Furthermore, we have shown the embedding diagram $z(r)$ by using the Eq. \eqref{6d2}, which is depicted in Fig. \ref{fig16}.\\
To analyze the energy conditions within the context of the Lorentzian distribution, we will employ the shape function from equation \eqref{34} in equations \eqref{19} and \eqref{20}. This yields the following pressure components: 
\begin{multline}\label{35}
\hspace{-0.3cm}p_r = \frac{{\alpha K_2}^2}{\mathcalboondox{L}_1}\left(\frac{216 \pi ^2-30 \pi  \beta +\beta ^2}{\beta +3 \pi }-\frac{2496 \beta  \sqrt{\Theta } M r^2}{\alpha  (\beta +48 \pi ) \left(\Theta +r^2\right)^2}
\right. \\ \left.
 +\frac{\left(312 \beta ^2 \sqrt{\Theta } M r^2\right)}{\pi  \alpha  (\beta +48 \pi ) \left(\Theta +r^2\right)^2}+\frac{78 \beta ^3 \sqrt{\Theta } M r^2}{\pi ^2 \alpha  (\beta +48 \pi ) \left(\Theta +r^2\right)^2}
 \right. \\ \left.
 -24 \pi +2 \beta-\frac{\sqrt{\Theta}}{(\beta +3 \pi ) \left({r_0}^2+\Theta \right)\mathcalboondox{M}_3}\left(27 (4 \pi -\beta )
  \right.\right. \\ \left.\left.
\times (\beta +8 \pi ) {r_0}^{2\mathcalboondox{L}_3} r^{-2\mathcalboondox{L}_3} \left(\pi ^2 \alpha  {r_0}^2 (12 \pi -\beta ) (\beta +48 \pi )
\right.\right.\right. \\ \left.\left.\left.
 +\pi ^2 \alpha  (12 \pi -\beta ) (\beta +48 \pi ) \Theta +24 (4 \pi -\beta ) (\beta +3 \pi ) 
 \right.\right.\right. \\ \left.\left.\left.
\times (\beta +8 \pi ) \sqrt{\Theta } M\right)\right)+\frac{1}{\mathcalboondox{M}_3}\left(648 (\beta -4 \pi )^2 (\beta +8 \pi )^2
  \right.\right. \\ \left.\left.
\times M \left({r_0}^{2\mathcalboondox{L}_3} r^{-2\mathcalboondox{L}_3} \, _2F_1\left(1,\mathcalboondox{L}_3,\mathcalboondox{L}_4,-\frac{{r_0}^2}{\Theta }\right)+\frac{\Theta }{\Theta +r^2}
\right.\right.\right. \\ \left.\left.\left.
-\,_2F_1\left(1,\mathcalboondox{L}_3,\mathcalboondox{L}_4,-\frac{r^2}{\Theta }\right)\right)\right)\right)\,,
\end{multline}
\begin{equation}\label{36}
p_t = \frac{\alpha }{(\beta +8 \pi ) r^2}-\frac{\sqrt{\Theta } M}{\pi ^2 \left(\Theta +r^2\right)^2}\,,
\end{equation}
where $\mathcalboondox{M}_3= \pi ^2 \alpha  (\beta +48 \pi )^2 \sqrt{\Theta }$.\\
Also, at the wormhole throat $r=r_0$, NEC and SEC are as follows
\begin{equation}\label{3211}
\rho + p_r \bigg\vert_{r=r_0}=\frac{4}{\beta +48 \pi} \left(\frac{\alpha  (\beta -12 \pi )}{{r_0}^2 (\beta +8 \pi )}+\frac{3 (4 \pi -\beta ) \sqrt{\Theta } M}{\pi ^2 \left({r_0}^2+\Theta \right)^2}\right)\,,
\end{equation}
\begin{equation}
\rho + p_r + 2 p_t \bigg\vert_{r=r_0}= \frac{2}{\beta +48 \pi} \left(\frac{3 \alpha }{a^2}-\frac{(7 \beta +24 \pi ) \sqrt{\Theta } M}{\pi ^2 \left(a^2+\Theta \right)^2}\right) \,.
\end{equation}
After thoroughly examining the given expressions, it becomes clear that $\beta$ cannot assume values of $-8\pi$ or $-48\pi$. Considering this constraint, we generated energy condition graphs, depicted in Figs. \ref{fig5} and \ref{fig6}. The left graph of Fig. \ref{fig5} presents the energy density as a radial coordinate $r$ function, displaying a consistent decrease across the entire space-time. In contrast, the right graph in the same figure illustrates the behavior of the SEC for various $\beta$ values, consistently indicating a negative trend. Furthermore, from Fig. \ref{fig6}, the left graph demonstrates the radial NEC's adverse behavior, signaling an NEC violation. Conversely, the right graph in Fig. \ref{fig6} confirms the favorable behavior of the tangential NEC, which diminishes as we move away from the wormhole throat.\\
 \begin{figure*}
\centering
\includegraphics[width=13.5cm,height=5.5cm]{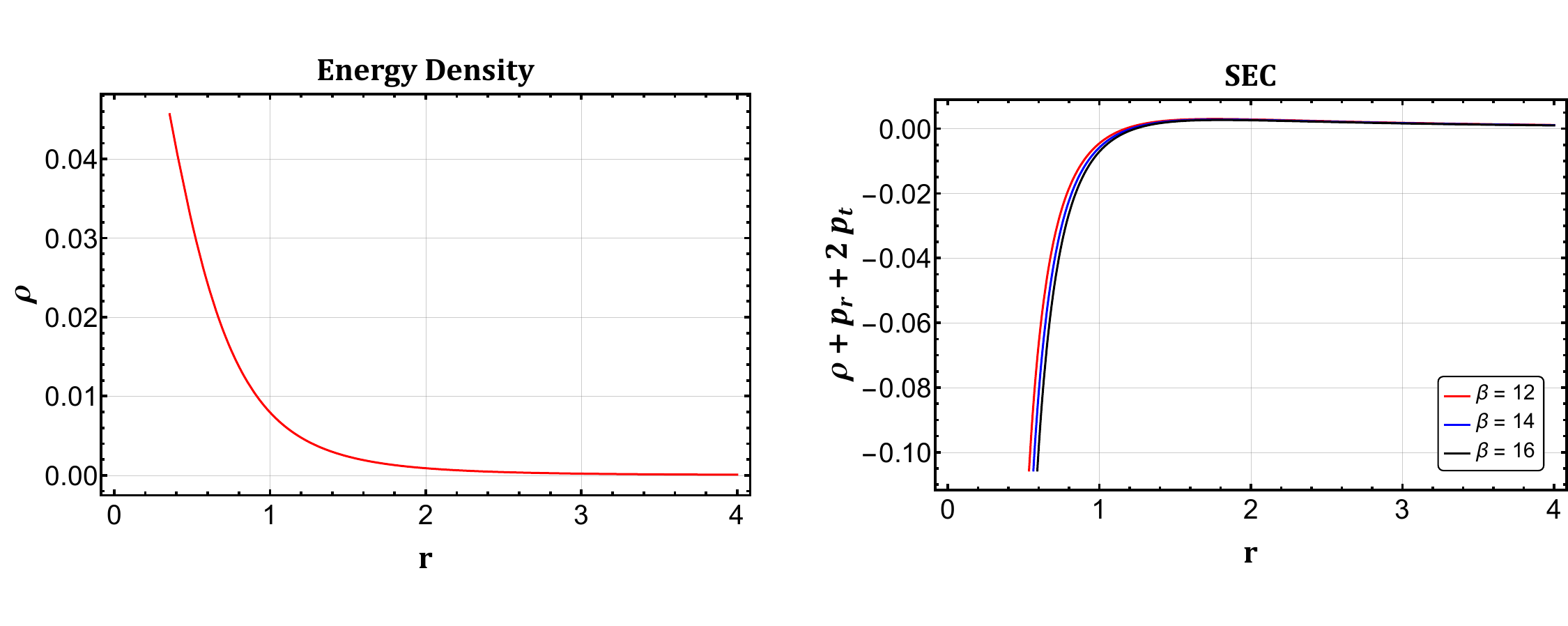}
\caption{The figure displays the variations in the energy density \textit{(on the left)} and the SEC \textit{(on the right)} as a function of the radial coordinate `$r$ ' for various values of `$\beta$ '. Additionally, we maintain fixed values for other parameters, including $\alpha=0.3,\, \Theta=0.5,\, M=0.25,\, K_2=0.2,\, \text{and} \, r_0 = 1$.}
\label{fig5}
\end{figure*}
 \begin{figure*}
\centering
\includegraphics[width=14.5cm,height=5cm]{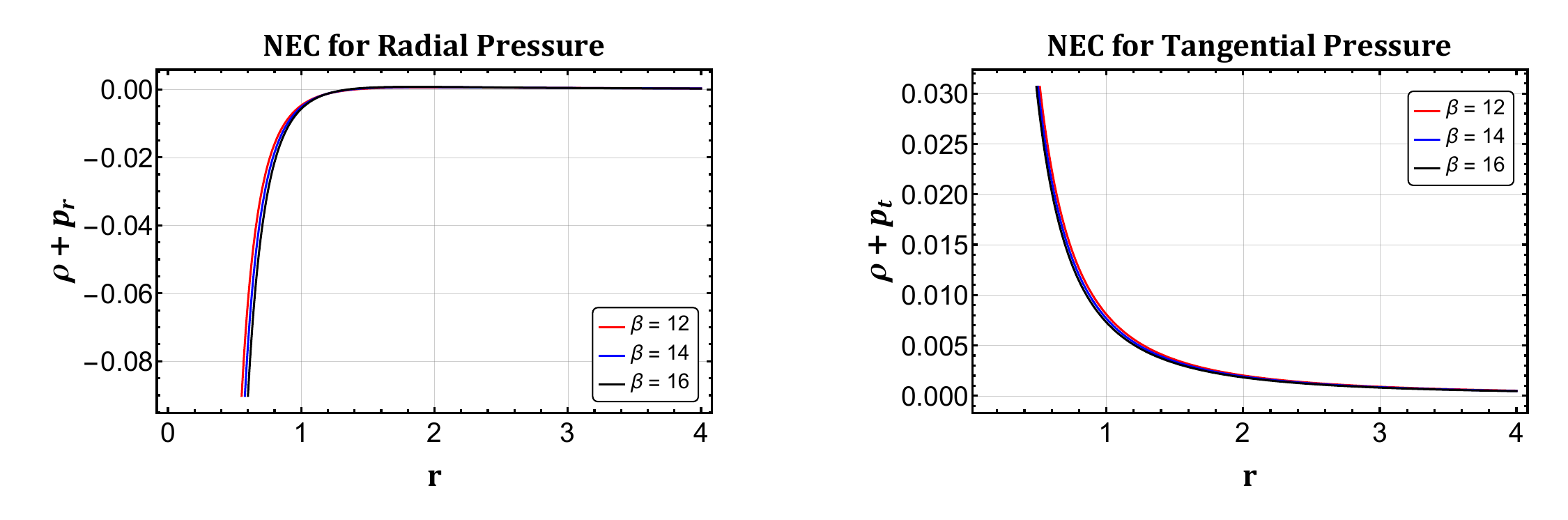}
\caption{The figure displays the variations in the NEC for radial \textit{(on the left)} and tangential \textit{(on the right)} pressure as a function of the radial coordinate `$r$ ' for various values of `$\beta$ '. Additionally, we maintain fixed values for other parameters, including $\alpha=0.3,\, \Theta=0.5,\, M=0.25,\, K_2=0.2,\, \text{and} \, r_0 = 1$.}
\label{fig6}
\end{figure*}
In the next Section \ref{sec5}, we will estimate the amount of exotic matter using the volume integral quantifier parameter.

\section{Amount of exotic matter}\label{sec5}
\begin{figure*}
\centering
\includegraphics[width=14.5cm,height=5.5cm]{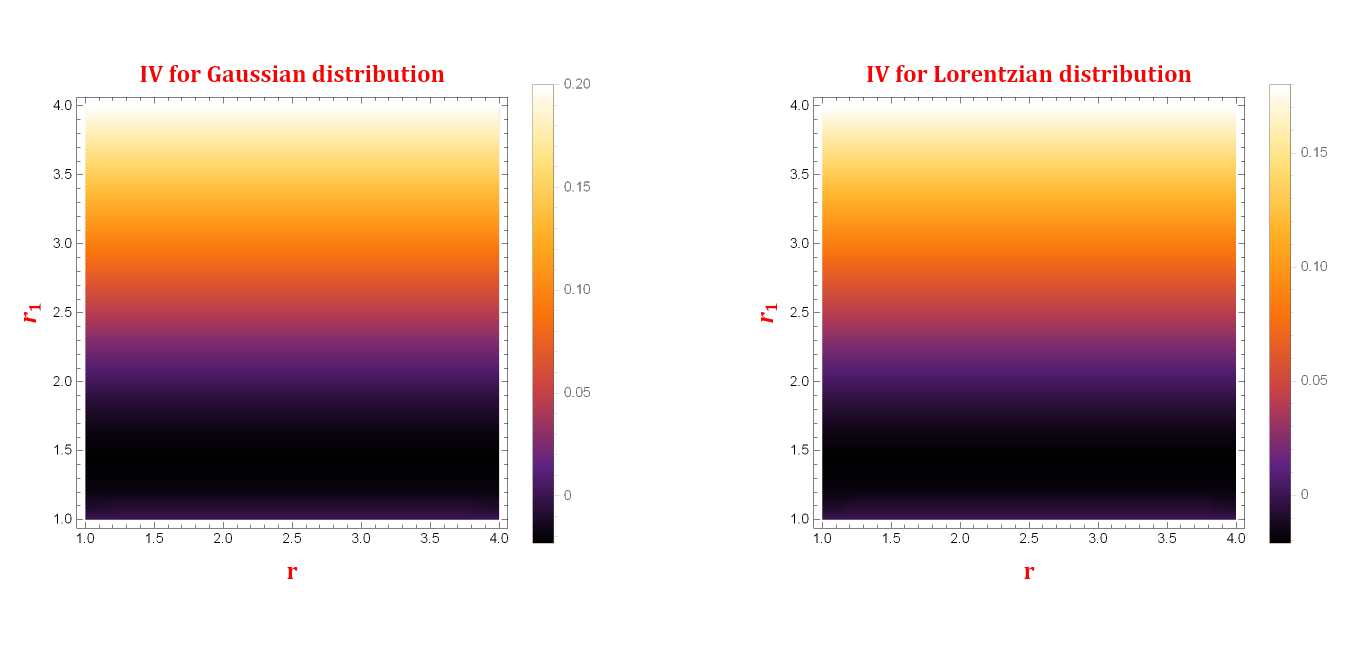}
\caption{The figure displays the variations in the VIQ. Additionally, we maintain fixed values for other parameters, including $\alpha=0.3,\, \Theta=0.5,\, M=0.25,\, K_2=0.2,\, \text{and} \, r_0 = 1$.}
\label{fig11}
\end{figure*}

Now, let us proceed with estimating the quantity of exotic matter necessary for ensuring the stability of a wormhole. To achieve this, we employ the Volume Integral Quantifier (VIQ) method, as introduced by Visser et al. in their work \cite{M. Visser}, which offers a means to quantify the average quantity of matter within space-time that violates the Null Energy Condition (NEC). The VIQ can be formally defined as follows:
\begin{equation}\label{37}
IV=\oint \left[\rho+P_r\right]dV\,.
\end{equation}
The volume can be interpreted as $dV=r^2\,dr\,d\Omega$ with $d\Omega$ the solid angle. Since $\oint dV=2\int_{r_0}^{\infty}dV=8\pi \int_{r_0}^{\infty}r^2dr,$ we have
\begin{equation}\label{38}
IV=8\pi \int_{r_0}^{\infty}(\rho+P_r)r^2dr.
\end{equation}
As seen in the equation above, the integration bounds extend to infinity. As previously established in reference \cite{F.S.N. Lobo}, for a wormhole to exhibit asymptotically flat characteristics, the Volume Integral Quantifier (VIQ) with infinite bounds over the radial coordinate must exhibit divergence. Consequently, it would be beneficial to introduce an energy-momentum tensor cut-off scale at a specific location denoted as $r_1$. Next, we can express the volume integral for a wormhole characterized by a field that varies from the throat at $r_0$ to a specified radius $r_1$, with the condition that $r_1\geq r_0,$. This volume integral is defined as follows
\begin{equation}\label{1122}
IV=8\pi \int_{r_0}^{r_1}(\rho+P_r)r^2dr.
\end{equation}
Utilizing the equation \eqref{1122}, we have explored the volume integral and visualized its behavior in Fig. \ref{fig11}. The graphs clearly illustrate that as $r_1$ approaches $r_0$, the value of $IV$ tends towards zero. This observation leads us to conclude that a relatively small amount of exotic matter can stabilize a traversable wormhole. We have also determined that choosing an appropriate wormhole geometry can minimize the total quantity of matter violating the Average NEC. For further insights and intriguing applications of the VIQ, interested readers can refer to the references \cite{K. Jusufi, O. Sokoliuk}.

\section{Complexity factor}\label{sec6}
\begin{figure*}
\centering
\includegraphics[width=15.5cm,height=5cm]{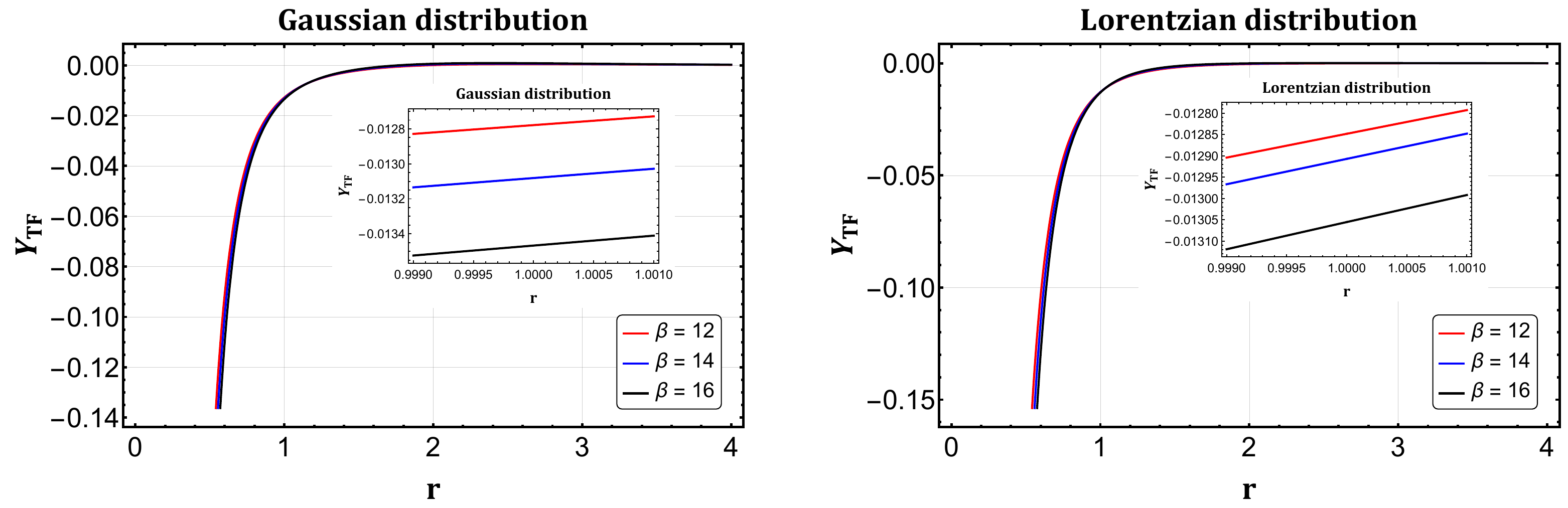}
\caption{The figure displays the variations in the complexity factor as a function of the radial coordinate `$r$ ' for various values of `$\beta$ '. Additionally, we maintain fixed values for other parameters, including $\alpha=0.3,\, \Theta=0.5,\, M=0.25,\, K_2=0.2,\, \text{and} \, r_0 = 1$.}
\label{fig12}
\end{figure*}

In 2018, Herrera presented the idea of the complexity factor for self-gravitating, spherically symmetric, static systems within the framework of GR \cite{L. Herrera}. The essence of the complexity factor is associated with systems that are isotropically pressured, have uniform energy density, and are either relatively simple or exhibit minimum intricacy. The complexity factor in these cases is found to be zero. Furthermore, the complexity factor does not change in the presence of anisotropic pressure or non-uniform energy density as long as their combined effects balance one another out. $Y_{TF}$, the traced free scalar complexity factor, has the following expression:
\begin{equation}\label{39}
Y_{TF}=p_r - p_t  -\dfrac{1}{2r^{3}}\int _{r_{0}}^{r}r^{3}\dfrac{d\rho (r)}{dr}dr\,. 
\end{equation}
By substituting Equations \eqref{21}, \eqref{a4}, and \eqref{a5} into the preceding equation, we can determine the complexity factor for the Gaussian distribution as given below
\begin{multline}\label{40}
Y_{TF}= \frac{1}{16 r^3}\left(-\frac{72 \alpha  (12 \pi -\beta ) {r_0}^{2\mathcalboondox{L}_3} r^{-\frac{15 \beta }{\beta +48 \pi }}}{(\beta +3 \pi ) (\beta +48 \pi )}
  \right. \\ \left.
  +\frac{r_0 M e^{-\frac{{r_0}^2}{4 \Theta }} \left({r_0}^2+6 \Theta \right)}{\pi ^{3/2} \Theta ^{3/2}}+\frac{24 r \alpha(2\pi-\beta)}{(\beta +3 \pi)(\beta +8 \pi)}+\frac{6 M }{\pi }
   \right. \\ \left.
  \times \left(\text{erf}\left(\frac{r}{2 \sqrt{\Theta }}\right)-\text{erf}\left(\frac{r_0}{2 \sqrt{\Theta }}\right)\right)-\frac{M r e^{-\frac{r^2}{4 \Theta }}\left(6 \Theta +r^2\right)}{\pi ^{3/2} \Theta ^{3/2}} 
   \right. \\ \left.
 \times -\frac{216 (4 \pi -\beta ) (\beta +8 \pi ) }{\pi ^{3/2} (\beta +48 \pi )^2 \Theta ^{3/2}}\left(M {r_0}^{2(\mathcalboondox{L}_3+1)} r^{-\frac{15 \beta }{\beta +48 \pi }} 
   \right.\right. \\ \left.\left.
 \times E_{-\mathcalboondox{L}_3}\left(\frac{{r_0}^2}{4 \Theta }\right)\right)+\frac{6 (4 \pi -\beta ) M r^5 E_{\mathcalboondox{L}_4}\left(\frac{r^2}{4 \Theta }\right)}{\pi ^{3/2} (\beta +48 \pi ) \Theta ^{5/2}}\right)\,.
\end{multline}
Here, the term ``erf" corresponds to the error function, which can be defined as $\text{erf}(Z)=\frac{2}{\sqrt{\pi}}\bigintsss\limits_{0}^{Z}e^{-t^2}dt$.
Additionally, by employing Equations \eqref{24}, \eqref{35}, and \eqref{36} within Equation \eqref{39}, we can establish the complexity factor for the Lorentzian distribution as follows:
\begin{multline}\label{41}
Y_{TF}=\frac{\sqrt{\Theta } M}{\pi ^2 \left(\Theta +r^2\right)^2}-\frac{\alpha }{(\beta +8 \pi ) r^2}+\frac{2 \sqrt{\Theta } M }{\pi ^2 r^3}\\
\times \left(\frac{5 {r_0}^3+3 {r_0} \Theta }{8 \left({r_0}^2+\Theta \right)^2}-\frac{3 \tan ^{-1}\left(\frac{{r_0}}{\sqrt{\Theta }}\right)}{8 \sqrt{\Theta }}-\frac{5 r^3+3 \Theta  r}{8 \left(\Theta +r^2\right)^2}
  \right. \\ \left.
+\frac{3 \tan ^{-1}\left(\frac{r}{\sqrt{\Theta }}\right)}{8 \sqrt{\Theta }}\right)+\frac{\alpha}{6 (4 \pi -\beta ) (\beta +8 \pi ) r^2}\\
\times \left(\frac{\beta ^2}{\beta +3 \pi }+2 \beta +\frac{216 \pi ^2}{\beta +3 \pi }-\frac{2496 \beta  \sqrt{\Theta } M r^2}{\alpha  (\beta +48 \pi ) \left(\Theta +r^2\right)^2}
 \right. \\ \left.
+\frac{78 \beta ^3 \sqrt{\Theta } M r^2}{\pi ^2 \alpha  (\beta +48 \pi ) \left(\Theta +r^2\right)^2}+\frac{312 \beta ^2 \sqrt{\Theta } M r^2}{\pi  \alpha  (\beta +48 \pi ) \left(\Theta +r^2\right)^2}
 \right. \\ \left.
-24 \pi-\frac{30 \pi  \beta }{\beta +3 \pi }-\frac{27 (4 \pi -\beta ) (\beta +8 \pi )  }{\pi ^2 \alpha  (\beta +3 \pi ) (\beta +48 \pi )^2 \left({r_0}^2+\Theta \right)}
\right. \\ \left.
\times \left(\pi ^2 \alpha  {r_0}^2 (12 \pi -\beta ) (\beta +48 \pi )+\pi ^2   (12 \pi -\beta ) (\beta +48 \pi ) 
\right.\right. \\ \left.\left.
\times \alpha\Theta +24 (4 \pi -\beta ) (\beta +3 \pi ) (\beta +8 \pi ) \sqrt{\Theta } M\right)
\right. \\ \left.
+\frac{648 (\beta -4 \pi )^2 (\beta +8 \pi )^2 M}{\pi ^2 \alpha  (\beta +48 \pi )^2 \sqrt{\Theta }}\left(\frac{\Theta }{\Theta +r^2}+{r_0}^{2\mathcalboondox{L}_3} r^{-2\mathcalboondox{L}_3} 
\right.\right. \\ \left.\left.
\times \, _2F_1\left(1,\mathcalboondox{L}_3,\mathcalboondox{L}_4,-\frac{{r_0}^2}{\Theta }\right)-\, _2F_1\left(1,\mathcalboondox{L}_3,\mathcalboondox{L}_4,-\frac{r^2}{\Theta }\right)\right)\right)\,.
\end{multline}
From both the Eqs. \eqref{40} and \eqref{41}, it is clear that one can not assume $\beta$ as $-3\pi$, $-8\pi$, and $-48\pi$. Also, in the vicinity of the wormhole throat, the complexity factor exhibits a continuously increasing behavior, while at greater radial distances, $Y_{TF}$ tends to zero. The behavior of the complexity factor over the radial coordinate $r$ for both the Gaussian and Lorentzian distributions is depicted in Figure \ref{fig12}. Notably, we observe $Y_{TF}$ approaching zero as $r$ tends towards infinity. Here, $Y_{TF}$ is the traced free scalar complexity factor acting on the wormhole throat. The physical meaning of this statement is that the complexity factor acting on the wormhole throat becomes negligible as the radial coordinate r tends to infinity. This implies that the wormhole throat becomes more stable as r tends to infinity, and it is less likely to collapse under the influence of complexity factor. Furthermore, within the dynamics of the complexity factor, it becomes evident that pressure isotropy exerts a more significant influence than the uniformity of energy density.

\section{Equilibrium conditions}\label{sec7}
In this current section, we explore the application of the generalized Tolman-Oppenheimer-Volkov (TOV) equation, as cited in references \cite{Oppenheimer, Gorini, Kuhfittig} to assess the stability of the wormhole solutions we have derived. The expression for the generalized TOV equation is as follows:
\begin{eqnarray}\label{51}
\frac{\varpi^{'}}{2}(\rho+p_r)+\frac{dp_r}{dr}+\frac{2}{r}(p_r-p_t)=0\,,
\end{eqnarray}
where $\varpi=2\phi(r)$.\\
Because of the anisotropic distribution of matter, we define the hydrostatic ($F_h$), gravitational ($F_g$), and anisotropic ($F_a$) forces as follows:
\begin{eqnarray}\label{52}
\hspace{-0.3cm} F_h=-\frac{dp_r}{dr}, ~F_g=-\frac{\varpi^{'}}{2}(\rho+p_r), ~F_a=\frac{2}{r}(p_t-p_r).
\end{eqnarray}
In order to establish equilibrium for the wormhole solutions, it is essential that the condition $F_h + F_g + F_a = 0$ is satisfied. 

 \begin{figure}[H]
\centering
\includegraphics[width=6cm,height=5cm]{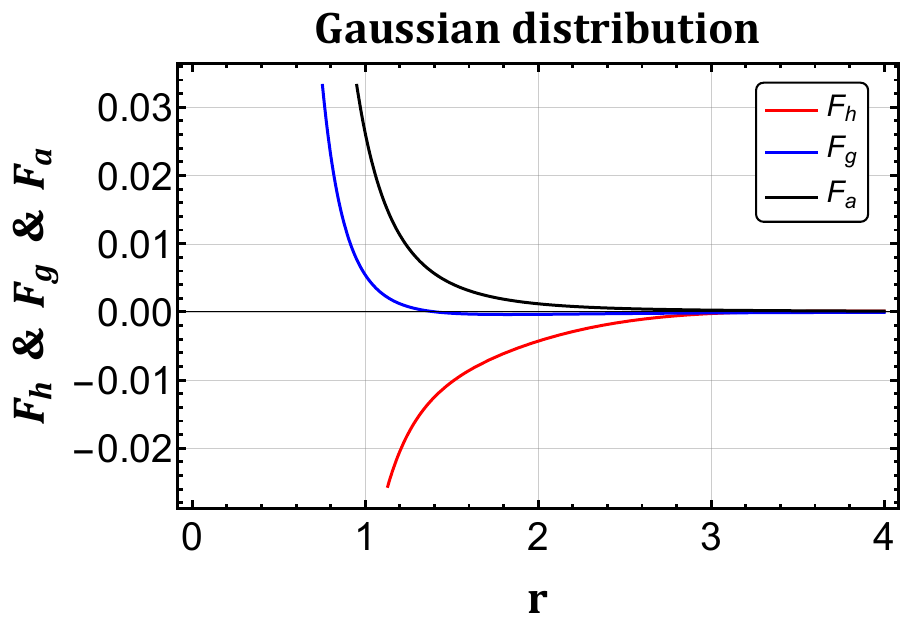}
\caption{The figure displays the variations of forces under Gaussian distribution as a function of the radial coordinate `$r$ ' for various values of `$\beta$ '. Additionally, we maintain fixed values for other parameters, including $\alpha=0.3,\, \Theta=0.5,\, M=0.25,\, K_2=0.2,\, \text{and} \, r_0 = 1$.}
\label{fig13}
\end{figure}

 \begin{figure}[H]
\centering
\includegraphics[width=6cm,height=5cm]{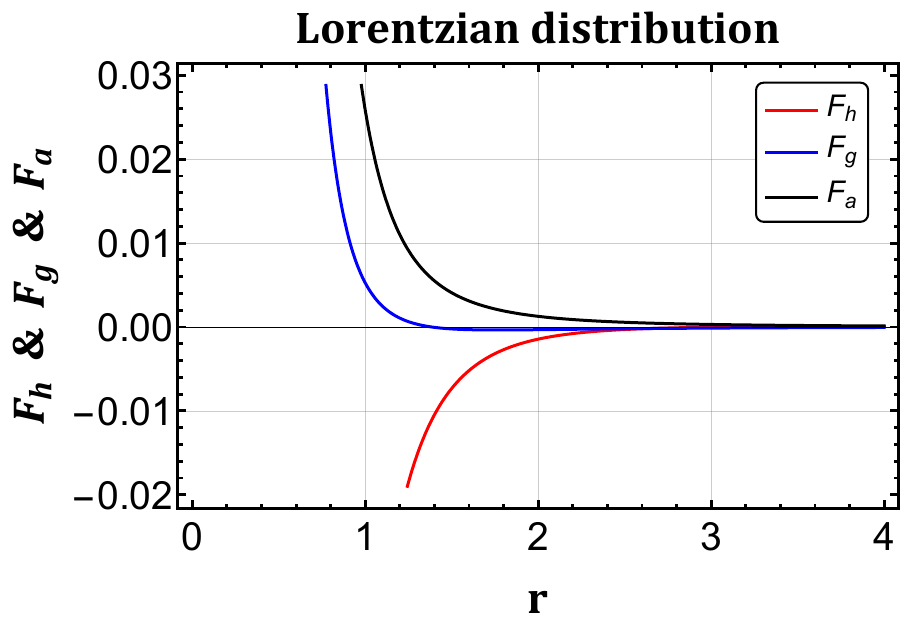}
\caption{The figure displays the variations of forces under Lorentzian distribution as a function of the radial coordinate `$r$ ' for various values of `$\beta$ '. Additionally, we maintain fixed values for other parameters, including $\alpha=0.3,\, \Theta=0.5,\, M=0.25,\, K_2=0.2,\, \text{and} \, r_0 = 1$.}
\label{fig14}
\end{figure}
In Figures \ref{fig13} and \ref{fig14}, we depict the profiles of hydrostatic, gravitational, and anisotropic forces for our wormhole solutions when Gaussian and Lorentzian sources are employed, utilizing the same parameter values as in the preceding figures. Notably, it is evident from these figures that the $F_h$ holds a dominant position in comparison to the $F_a$ and $F_g$ forces for both source scenarios. In both cases, $F_h$ exhibits positive values, while $F_a$ and $F_g$ assume negative values. This observation underscores that for the system to remain in equilibrium, the $F_h$ force must be counterbalanced by the combined influence of $F_a$ and $F_g$. Interested readers may refer to Refs for a more in-depth exploration of this topic. \cite{Farook Rahaman, Shamaila Rani}.

\section{Conclusions}
\label{sec8}
In this paper, we explore the potential existence of wormhole solutions within the framework of $f(Q, T)$ symmetric teleparallel gravity, considering anisotropic matter sources. The underlying geometry of this investigation draws inspiration from non-commutativity and the presence of conformal symmetry. It is well-established that a conformal killing vector field exists when space-time exhibits conformal symmetry. This presence of conformal symmetry plays a crucial role by reducing the number of unknown quantities involved in the analysis. Using CKVs and non-commutative sources for finding solutions is not a novel approach and has been explored in various modified theories. In \cite{Ghosh}, Ghosh examined the existence of an Einstein-Gauss-Bonnet black hole inspired by non-commutative geometry. Conformally symmetric wormhole solutions have been investigated under different equations of state (EoS) in modified teleparallel gravity \cite{Singh1}. Moreover, in the context of $f(Q)$ gravity, non-commutative distributed wormhole solutions with conformal symmetry have been discussed in \cite{Singh2}. However, our current investigation presents a novel contribution by employing non-commutative distributions and conformal symmetry in the framework of $f(Q, T)$, a combination not explored previously. Throughout this study, we adopt a specific form of the linear model for $f(Q)$, expressed as $f(Q) = \alpha Q + \beta T$, where $\alpha$ and $\beta$ serve as free parameters. Drawing inspiration from the works of Nicolini et al. \cite{P. Nicoloni} and Mehdipour \cite{Mehdipour11}, we incorporate the non-commutative energy density proposed by them. Subsequently, we analyze shape functions, energy conditions, and the stability of the wormhole solutions analytically as well as graphically. The results are summarized as follows.\\
We have obtained the shape function of the wormholes under both distributions and graphically depicted the behavior of the shape functions. It is observed that the shape function obeys the flare-out condition in an asymptotic background under both Gaussian and Lorentzian distributions. We detected certain constraints on $\beta$, i.e., $\beta>0$, which ensure the satisfaction of the flare-out condition. Our investigation has revealed the influence of the model parameter $\beta$ and the non-commutative parameter $\Theta$ on shape functions, with a minimal impact in the throat region. Additionally, we observed that an increase in the model parameter $\beta$ leads to a decrease in the shape function for both distributions.\\
Further, we checked energy conditions, especially energy density, NEC and SEC. Energy density shows positively decreasing behavior throughout the space-time. Radial NEC is disrespected at the throat ($r_0=1$) and its vicinity. Interestingly, we noticed that for higher values of $\beta$, NEC is disrespected (see  Eqs. \eqref{3210} and \eqref{3211}). Moreover, we plotted the graphs for SEC for both distributions and found that SEC is violated near the throat. The permission for the existence of the wormhole is encouraged by the violation of the NEC. Consequently, the derived solutions are considered viable, establishing the feasibility of the wormhole's presence within the framework of $f(Q, T)$ gravity under these specific non-commutative geometries.\\
Furthermore, the amount of exotic matter via VIQ has been investigated to assess the quantity of exotic matter needed at the wormhole throat to be traversable. Our analysis indicates that a relatively small amount of exotic matter is necessary to maintain traversability. This leads us to the conclusion that modifications to standard GR can effectively reduce the reliance on exotic matter, offering a stable and viable solution for a traversable wormhole.\\
Finally, the stability analysis of the solutions for both distributions has been investigated through the generalized TOV equation. The graphs show that hydrostatic force is counterbalanced by the combined influence of anisotropic force and gravitational force, which confirms the stable behavior of our solutions. Moreover, the complexity factor of the non-commutative wormhole has also been calculated, and the monotonically increasing behavior of the complexity factor has also been observed.\\
Note that in \cite{T1}, the authors discussed wormhole solutions under Gaussian and Lorentzian geometries for the linear as well as non-linear forms of $f(Q, T)$ gravity. They investigated wormhole solutions analytically for the linear case and numerically for the non-linear model. Moreover, they examined the gravitational lensing phenomenon for the exact wormhole solution and determined that at the wormhole throat, the deflection angle diverges. In another study,  Chalavadi et al. \cite{T2} examined wormhole solutions for traceless energy-momentum tensor and linear EoS relation for both non-commutative distributions in $f(Q, T)$ lagrangian. Notably, conformally symmetric wormholes under non-commutative geometry have yet to be discussed in this modified gravity, which motivates us to study this gap. Our overall theoretical observation is that in our current approach, the solutions and properties of the model are physically valid and exciting as much as in the former method.

\section*{Data Availability Statement}
There are no new data associated with this article.

\acknowledgments  MT acknowledges the University Grants Commission (UGC), New Delhi, India, for awarding the National Fellowship for Scheduled Caste Students (UGC-Ref. No.: 201610123801). ZH acknowledges the Department of Science and Technology (DST), Government of India, New Delhi, for awarding a Senior Research Fellowship (File No. DST/INSPIRE Fellowship/2019/IF190911). PKS acknowledges National Board for Higher Mathematics (NBHM) under the Department of Atomic Energy (DAE), Govt. of India, for financial support to carry out the Research project No.: 02011/3/2022 NBHM(R.P.)/R\&D II/2152 Dt.14.02.2022. We are very much grateful to the honorable referee and to the editor for the illuminating suggestions that have significantly improved our work in terms of research quality and presentation. 



\begin{thebibliography}{52}
\footnotesize
\bibitem{Einstein1} A. Einstein and N. Rosen, \textit{Phys. Rev.} \textbf{48}, 73 (1935).
\bibitem{Einstein2} J. A. Wheeler, \textit{Phys. Rev.} \textbf{97}, 511 (1955).
\bibitem{Einstein3} D. R. Brill and J. A. Wheeler, \textit{Rev. Mod. Phys.} \textbf{29}, 465 (1957).
\bibitem{Einstein4} D. R. Brill and J. B. Hartle, \textit{Phys. Rev.} \textbf{135}, B271 (1964).
\bibitem{Einstein5} R. W. Fuller and J. A. Wheeler, \textit{Phys. Rev.} \textbf{128}, 919 (1962).
\bibitem{Thorne/1988} M. S. Morris and K. S. Thorne, \textit{Am. J. Phys.} \textbf{56}, 395-412 (1988).
\bibitem{Thorne/1988a} M. S. Morris, K. S. Thorne and U. Yurtsever, \textit{Phys. Rev. Lett.} \textbf{61}, 1446 (1988).
\bibitem{Visser1} M. Visser, \textit{Lorentzian Wormholes: From Einstein to Hawking} (American Institute of Physics, New York, 1995).
\bibitem{Armendariz} C. Armendariz-Picon, \textit{Phys. Rev. D} \textbf{65}, 104010 (2002).
\bibitem{Lobo1} F. S. N. Lobo, \textit{Phys. Rev. D} \textbf{71}, 084011 (2005).
\bibitem{Sushkov} S. V. Sushkov, \textit{Phys. Rev. D} \textbf{71}, 043520 (2005).
\bibitem{Wheeler3} J. A. Wheeler, \textit{Geons. Phys. Rev.} \textbf{97}, 511 (1955).
\bibitem{Mak} T. Harko, F. S. N. Lobo, M. K. Mak, and S.V. Sushkov,\textit{ Phys. Rev. D} \textbf{87}, 067504 (2013).
\bibitem{Zangeneh} M. K. Zangeneh, F. S. N. Lobo, N. Riazi, \textit{Phys. Rev. D} \textbf{90}, 024072 (2014).
\bibitem{Galiakhmetov} K. A. Bronnikov, A. M. Galiakhmetov, \textit{Gravity Cosmol.} \textbf{21}, 283 (2015).
\bibitem{Kar2} R. Shaikh, S. Kar, \textit{Phys. Rev. D} \textbf{94}, 024011 (2016).
\bibitem{Ziaie} M.R. Mehdizadeh, A.H. Ziaie, \textit{Phys. Rev. D} \textbf{99}, 064033 (2019).
\bibitem{Singleton} V. D. Dzhunushaliev, D. Singleton, \textit{Phys. Rev. D} \textbf{59}, 064018 (1999).
\bibitem{Leon} J. P. de Leon, \textit{J. Cosmol. Astropart. Phys.} \textbf{0911}, 013 (2009).
\bibitem{Folomeev} V. Dzhunushaliev, V. Folomeev, \textit{Mod. Phys. Lett. A} \textbf{29}, 1450025 (2014).
\bibitem{Pavlovic} S. Bahamonde, M. Jamil, P. Pavlovic, M. Sossich, \textit{Phys. Rev. D} \textbf{94}, 044041 (2016).
\bibitem{Idris} F. Rahaman, A. Banerjee, M. Jamil, A. K. Yadav, H. Idris, \textit{Int. J. Theor. Phys.} \textbf{53}, 1910 (2014).
\bibitem{Halilsoy} S. H. Mazharimousavi, M. Halilsoy, \textit{Mod. Phys. Lett. A} \textbf{31}, 1650203 (2016).
\bibitem{Karakasis} T. Karakasis, E. Papantonopoulos, and C. Vlachos, \textit{Phys. Rev. D} \textbf{105}, 024006 (2022).
\bibitem{Golchin} H. Golchin, M. R. Mehdizadeh, \textit{Eur. Phys. J. C}\textbf{ 79}, 777 (2019).
\bibitem{Eid} A. Eid, \textit{Phys. Dark Univ.} \textbf{30}, 100705 (2020).
\bibitem{Goswami} D. J. Gogoi and U. D. Goswami, \textit{J. Cosmol. Astropart. Phys.} \textbf{2023}, 027 (2023).
\bibitem{Malik}  A. Malik, et al. \textit{Eur. Phys. J. C} \textbf{83}, 522 (2023).
\bibitem{Ahmad} M. Zubair, S. Waheed, Y. Ahmad, \textit{Eur. Phys. J. C} \textbf{76}, 444 (2016).
\bibitem{Bhatti} Z. Yousaf, M. Ilyas,  and M. Z. Bhatti, \textit{Eur. Phys. J. Plus} \textbf{132}, 268 (2017).
\bibitem{Chanda} A. Chanda, S. Dey, and B. C. Paul, \textit{Gen. Relativ. Grav.} \textbf{53}, 78 (2021).
\bibitem{Rosa} J. L. Rosa, P. M. Kull, \textit{Eur. Phys. J. C} \textbf{82}, 1154 (2022).
\bibitem{Tayde 1} M. Tayde, S. Ghosh, and P. K. Sahoo, \textit{Chinese Phys. C} {\bf 47}, 075102 (2023).
\bibitem{Tayde 2} M. Tayde, J. R. L. Santos, J. N. Araujo, and P. K. Sahoo, \textit{Eur. Phys. J. Plus} \textbf{138}, 539 (2023). 
\bibitem{Boehmer} C. G. Boehmer, T. Harko, F. S. N. Lobo, \textit{Phys. Rev. D} \textbf{85}, 044033 (2012).
\bibitem{Rani} M. Sharif and Shamaila Rani, \textit{Phys. Rev. D} \textbf{88}, 123501 (2013).
\bibitem{Momeni} M. Jamil, D. Momeni, R. Myrzakulov, \textit{Eur. Phys. J. C} \textbf{73}, 2267 (2013).

\bibitem{Y.Xu}  Y. Xu et al., \textit{Eur. Phys. J. C} \textbf{79}, 708 (2019).
\bibitem{Arora111} A. Najera and A. Fajardo,, \textit{Phys. Dark Univ.} \textbf{34}, 100889 (2021).
\bibitem{Shiravand} M. Shiravand, S. Fakhry, and M. Farhoudi, \textit{Phys. Dark Univ.} \textbf{37}, 101106 (2022).
\bibitem{Bhattacharjee} S. Bhattacharjee and P. K. Sahoo, \textit{Eur. Phys. J. C} \textbf{80}, 289 (2020).
\bibitem{Najera} A. Najera and A. Fajardo, \textit{J. Cosmo. Astro. Phys.} \textbf{2022}, 020 (2022).
\bibitem{Gadbail} G. N. Gadbail, S. Arora, and P. K. Sahoo, Phys. Lett. B \textbf{838}, 137710 (2023).
\bibitem{Tayde12}  M. Tayde, Z. Hassan, P. K. Sahoo, and S. Gutti, \textit{Chinese Phys. C} \textbf{46}, 115101 (2022).
\bibitem{Bourakadi1} K. El Bourakadi, M. Koussour, G. Otalora, M. Bennai, and T. Ouali, \textit{Phys. Dark Univ.} \textbf{41}, 101246 (2023).
\bibitem{Smailagic1} P. Nicolini, A. Smailagic, E. Spallucci, \textit{Phys. Lett. B} \textbf{632},547-551 (2006).
\bibitem{Snyder1} H. S. Snyder, \textit{Phys. Rev.} \textbf{71}, 38 (1947).
\bibitem{Snyder2} H. S. Snyder, \textit{Phys. Rev.} \textbf{72}, 68 (1947).
\bibitem{Smailagic2} Mathew Schneider and Andrew DeBenedictis, \textit{Phys. Rev. D} \textbf{102}, 024030 (2020).
\bibitem{Stephan} C. A. Stephan, \textit{arXiv} arXiv:1305.3066 (2013).
\bibitem{Witten1} E. Witten, \textit{Nucl. Phys. B} \textbf{460}, 335 (1996).
\bibitem{Witten2} N. Seiberg, E. Witten, \textit{J. High Ener. Phys.} \textbf{9909}, 032 (1999).
\bibitem{Spallucci3}  A. Smailagic and E. Spallucci, \textit{J. Phys. A} \textbf{36}, L467 (2003).
\bibitem{Nozari1} K. Nozari and S. H. Mehdipour, \textit{J. High Energy Phys.} \textbf{2009}, 061 (2009).
\bibitem{Nozari2} S. Sushkov, \textit{Phys Rev. D } \textbf{71}, 043520 (2005).
\bibitem{Nozari3} F. Rahaman, S. Islam, P. K. F. Kuhfittig, and S. Ray, \textit{Phys. Rev. D} \textbf{86}, 106010 (2012).
\bibitem{Nozari4} F. Rahaman, et al. \textit{Phys. Rev. D} \textbf{87}, 084014 (2013).
\bibitem{Rani1} M. Sharif and S. Rani, \textit{Phys. Rev. D} \textbf{88}, 123501 (2003).
\bibitem{Rani2} Z. Hassan, G. Mustafa, and P. K. Sahoo, \textit{Symmetry} \textbf{13}, 1260 (2021).

\bibitem{Hall} G. S. Hall, ``Symmetries and Curvature Structure in General Relativity" \textit{World Scientific}, Singapore (2004).
%






\bibitem{Herrera1} L. Herrera, J. Jimenez, L. Leal, J. Ponce de Leon, M.
Esculpi, and V. Galina, \textit{J. Math. Phys.} \textbf{25}, 3274 (1984).
\bibitem{Herrera2} R. Maartens and M. S. Maharaj, \textit{J. Math. Phys.} \textbf{31},
151 (1990).
\bibitem{Herrera3} P. K. F. Kuhfittig, \textit{Ann. Phys.} \textbf{355}, 115 (2015).
\bibitem{Singh2} G. Mustafa et al. \textit{Ann. Phys.} \textbf{437}, 168751 (2022).
\bibitem{Jimenez}  J. B. Jimenez, L. Heisenberg and T. Koivisto, \textit{Phys. Rev. D}, \textbf{98}, 044048 (2018).
\bibitem{Correa} P. H. R. S. Moraes, R. A. C. Correa, and R. V. Lobato, \textit{J. Cos. Astro. Phys.} \textbf{07}, 029 (2017).
\bibitem{Arora11} S. Arora \textit{et al.}, \textit{Phys. Dark Univ.} \textbf{30}, 100664 (2020).

\bibitem{Loo1} Tee-How Loo, M. Koussour, and Avik De, \textit{Ann. Phys.} \textbf{454}, 169333 (2023).

\bibitem{P. Nicoloni} P. Nicoloni, A. Smailagic, and E. Spalluci, \textit{Phys. Lett. B} \textbf{632}, 547-551 (2006).

\bibitem{A. Smailagic1} A. Smailagic and E. Spalluci, \textit{J. Phys. A Math. Gen.} \textbf{36}, L467 (2003).

\bibitem{Mehdipour11} S. H. Mehdipour, \textit{Eur. Phys. J. Plus} \textbf{127}, 80 (2012).

\bibitem{M. Visser} M. Visser, S. Kar, N. Dadhich, \textit{Phys. Rev. Lett.} \textbf{90}, 201102 (2003).

\bibitem{F.S.N. Lobo} F. S. N. Lobo, F. Parsaei, N. Riazi, \textit{Phys. Rev. D} \textbf{87}, 084030 (2013).

\bibitem{K. Jusufi} K. Jusufi, P. Channuie, M. Jamil, \textit{Eur. Phys. J. C} \textbf{80}, 127 (2020).

\bibitem{O. Sokoliuk} O. Sokoliuk, S. Mandal, P.K. Sahoo, A. Baransky, \textit{Eur. Phys. J. C} \textbf{82}, 280 (2022).

\bibitem{L. Herrera} L. Herrera, \textit{Phys. Rev. D} \textbf{97}, 044010 (2018).

\bibitem{Oppenheimer} J. R. Oppenheimer and G. M. Volkoff, \textit{Phys. Rev.} \textbf{55}, 374 (1939).

\bibitem{Gorini} V. Gorini, U. Moschella, A. Y.  Kamenshchik, V. Pasquier and A. A. Starobinsky, \textit{Phys. Rev. D} \textbf{78}, 064064 (2008).

\bibitem{Kuhfittig} P. K. F. Kuhfittig, \textit{Fundamental J. Mod. Phys.} \textbf{14}, 23-31 (2020).

\bibitem{Farook Rahaman} Farook Rahaman, Sreya Karmakar, Indrani Karar, and Saibal Ray, \textit{Lett. B} \textbf{746}, 73 (2015).

\bibitem{Shamaila Rani} Shamaila Rani and Abdul Jawad, \textit{Adv. High Energy Phys.} \textbf{2016}, 7815242 (2016).



\bibitem{Ghosh} S. G. Ghosh, \textit{Class. Quant. Grav.} \textbf{35}, 085008 (2018).
\bibitem{Singh1} K. N. Singh et al. \textit{Phys. Rev. D} \textbf{101}, 084012 (2020).
\bibitem{T1} M. Tayde, Z. Hassan and P. K. Sahoo, \textit{Phys. Dark Univ.} \textbf{42}, 101288 (2023).
\bibitem{T2} C. C. Chalavadi , N. S. Kavya and V. Venkatesha, \textit{Eur. Phys. J. Plus} \textbf{138}, 885 (2023).

\end{thebibliography}
\end{document}